\documentclass[a4paper,12pt]{article}
\usepackage[dvips]{epsfig}
\font\tenmsa=msam10
\font\sevenmsa=msam7
\font\fivemsa=msam5
\newfam\msafam
\textfont\msafam=\tenmsa  \scriptfont\msafam=\sevenmsa
  \scriptscriptfont\msafam=\fivemsa
\def\hexnumber@#1{\ifcase#1 0\or1\or2\or3\or4\or5\or6\or7\or8\or9\or
        A\or B\or C\or D\or E\or F\fi }
\edef\msa@{\hexnumber@\msafam}
\mathchardef\lesssim="3\msa@2E
\mathchardef\gtrsim="3\msa@26

\begin{document}
\begin{titlepage}
\noindent
\begin{center}
{\bf AN INVESTIGATION OF THE  2D ATTRACTIVE HUBBARD MODEL}\\
\vspace{1cm}
R. Lacaze$^{ab}$\footnote{lacaze@spht.saclay.cea.fr},
A. Morel$^a$\footnote{morel@spht.saclay.cea.fr},
B. Petersson$^c$\footnote{bengt@physik.uni-bielefeld.de}
and J. Schr\"oper$^c$\footnote{joergs@physik.uni-bielefeld.de}\\
\vspace{0.6cm}
$^a$ Service de Physique Th\'eorique, CEA-Saclay, 
91191 Gif-sur-Yvette Cedex, France\\
$^b$ASCI, Bat. 506, Universit\'e Paris Sud, 91405 Orsay Cedex, France \\
$^c$ Fakult\"at f\"ur Physik, Universit\"at Bielefeld, 
Postfach 10 01 31, D-33501 Bielefeld, Germany\\
\vspace{1cm}
{\bf Abstract}
\end{center}

We present an investigation of the 2D attractive Hubbard model, considered as
an effective model relevant to superconductivity in  strongly interacting
electron systems. We use both hybrid Monte Carlo simulations and existing
hopping parameter expansions to explore the low temperature domain. The increase
of the static S-wave pair correlation with decreasing temperature is analyzed
in terms of an expected Kosterlitz-Thouless superconducting transition.
The evidence for this transition is weak: If it exists, its temperature is very
low and depends weakly on the band filling near half filling. 
The number of unpaired electrons remains nearly constant with temperature
at fixed attractive potential strength.  In contrast, the static magnetic
susceptibility decreases fast with temperature, and cannot be related only
to pair formation. We introduce a method by which the Pad\'e approximants of
the existing series for the susceptibility give sensible results down to rather
low temperature region, as shown by comparison with our numerical data.
\vfill
PACS 71.10.Fd 74.25.Dw 75.40.Mg
\vfill
\par\noindent
October 1996 \hskip 2 truecm Submitted for publication to
{\sl J. Phys. I (France)}
\par\noindent
BI-TP 96/05\\
T96/ 114\\
\end{titlepage}
%
%
\section*{1. Introduction}

The study of models of strongly interacting electrons is very important for
the understanding of high temperature superconductivity. The fundamental
mechanism of this phenomenon has not yet been clearly identified, and it
may be interesting to study effective models, which have strong local
attraction between the electrons. The two dimensional attractive Hubbard
\cite{Hub} model is an example of such a model whose phase diagram as well as
physical properties in the normal non superconducting phase can be compared
to properties of actual superconducting materials of the high $T_c$
superconductivity class.

The two dimensional attractive Hubbard model is a conceptually
simple model, which at low temperatures is expected to have a
Kosterlitz-Thouless transition \cite{KT} into a superconducting phase,
away from half filling. At half filling the model has further symmetries,
which prevents such a transition. In the absence of magnetic field, the
properties of the model depend on two parameters (apart from
the temperature) namely the coupling constant of the attractive local
interaction $U$ and the chemical potential $\mu$, measured in units of the
coefficient $k$ of the hopping term. The model is solvable in the free case
$U=0$ and the atomic limit $k=0$. In the interesting case where both $U$
and $k$ are different from zero the model can only be studied through
developments around the two solvable cases, high temperature series and
numerical simulation techniques. In contrast to the repulsive Hubbard model,
with the opposite sign of U, the fermionic determinant is positive also for
chemical potential different from zero. Thus the numerical simulations do not
suffer from the sign problem in this case. In fact, the two models are related
by a change of sign in $U$ and the exchange of chemical potential and magnetic
field.

In this paper, using a hybrid Monte Carlo algorithm, we perform a detailed
numerical study of the model in a wide range of values of the parameters. The
purpose is first to examine in detail the evidence for a Kosterlitz-Thouless
transition and, if it takes place, to estimate the critical temperature.
This can be done by studying the s-wave pair-field correlation function,
which should diverge at the transition on an infinite lattice.
Another interesting quantity is the magnetic susceptibility, sensitive to the
 presence of single electrons as contrast to those bound in pairs.
One may expect the susceptibility to disappear in the strong coupling limit,
where more and more pairs are formed. However, at fixed $U$ the temperature
behavior of the probability that a site has a single electron and the behavior
of the susceptibility are not the same.  Therefore the behavior in temperature
of the susceptibility of the remnant unpaired electrons is interesting to
investigate.  For this latter quantity we make use of the series expansion
in $k$ given in Ref.\cite{HOA}, and with the help of the Pad\'e approximants
method, extrapolate the series to low temperatures. We compare the results
with our numerical data and show that, at least at small and intermediate
couplings, rather low temperature may be reached analytically. If true also
in the repulsive case of the Hubbard model, where numerical simulations
suffer from the sign problem, series expansion might be interesting to use.
Unfortunately, no such series are available for the pair field correlator.
    
There have been some earlier numerical investigations of this model
\cite{Scal,Morscal}, leading in particular to estimates of the transition
temperature. We compare our estimates to theirs, when there is overlap in the
parameters. Our estimate of the transition temperature is considerably lower
than in these investigations. In particular, for the range of temperatures we
simulated, we do not see a really significant difference between the half
filled case, where no transition is expected, and away from half-filling.

In Section 2 we define the model and the observables, which we measure.
Section 3 is devoted to the formulation of the path integral formalism
needed for the numerical simulation. In Section 4 we discuss the algorithm
employed, which we have chosen as the Hybrid Monte Carlo algorithm.
In Section 5 we discuss our data on the pair field correlation, which
is a direct indicator of a transition into a superconducting phase.
In Section 6 we present results for the susceptibility, comparing our
numerical results with analytic results extrapolated from the series expansion.
Section 7, finally, contains a summary and our conclusions.

%
%
\section*{2. The Model}
%
%
The model is defined by the Hubbard Hamiltonian 
\begin{eqnarray} \label{hubham}
H&=&- k \sum_{\left< x,y \right>}
    \left( a^{\dagger}_{x} a_{y} +  a^{\dagger}_{y} a_{x} 
    +  b^{\dagger}_{x} b_{y} +  b^{\dagger}_{y} b_{x} \right)
   - U \sum_x \left( a^{\dagger}_{x} a_{x}
             -\frac{1}{2} \right)
             \left( b^{\dagger}_x b_{x}
               -\frac{1}{2} \right)
\nonumber       \\
 &&  - \mu \sum_{ x } \left( a^{\dagger}_{x} a_{x} +
                             b^{\dagger}_{x} b_{x} \right)
\hspace{12pt},
\end{eqnarray}
where x and y are sites on a two dimensional $N_s \times N_s$ square lattice
with $V = N_s^2$ sites, $a_{x}$ and $b_{x}$ are coordinate space annihilation
operators for spin-up and spin-down electrons respectively, and $k$ is the
nearest neighbor hopping parameter.  The coupling $U>0$ denotes the strength
of the attractive local interaction and $\mu$ is the chemical potential,
defined so that $\mu = 0$ at half filling, i.e. where the total particle number 
$\left< N \right> = V$.  Here
\begin{equation} \label{n}
N= \sum_{x} ( a^{\dagger}_{x} a_{x} + b^{\dagger}_{x} b_{x})
\hspace{12pt}.
\end{equation}
We also define the particle number density operator
\begin{equation}  \label{defn}
n = \frac{1}{V} N 
\hspace{12pt}.
\end{equation}

We do not couple the system to an external magnetic field.
Thus the model is invariant under an $SU(2)_{\rm spin}$.
For $\mu=0$ it is also invariant under another $SU(2)$ group \cite{YZ},
generated by the operators
\begin{eqnarray} \label{sup}
J^{\prime}_{-}&=&\sum_{x} (-)^x a_x b_x    \hspace{12pt},
\nonumber         \\
J^{\prime}_{+}&=&J^{\prime \dagger}_{-}    \hspace{12pt},
\nonumber         \\
J^{\prime}_{0}&=&\frac{1}{2} \left( N - V \right)  \hspace{12pt}.
\end{eqnarray}

The thermodynamics of the model is given by the partition function
\begin{equation}\label{part}
Z=\mbox{Tr}\left (e^{-\beta H} \right)
\hspace{12pt}.
\end{equation}

In the simulation we will measure average values of operators in this ensemble,
\begin{equation}
\left< O \right> = \mbox{Tr}(O e^{-\beta H })/Z
\hspace{12pt}.
\end{equation}

The basic equal time correlation functions, which give
information about the properties of the model are
the S-wave on site pairing correlation function
\begin{equation} \label{defp}
P(x-y)= \left< (a^{\dagger}_x b^{\dagger}_x + b_x a_x)(a^{\dagger}_y \
b^{\dagger}_y + b_ya_y)\right>
\hspace{12pt},
\end{equation}
the magnetization density correlation function, or magnetic susceptibility
\begin{equation} \label{defm}
\chi (x-y)= \beta \left<  (a^{\dagger}_x a_x - b^{\dagger}_x b_x) 
(a^{\dagger}_y
a_y - b^{\dagger}_yb_y) \right>
\hspace{12pt},
\end{equation}
and the charge density correlation function
\begin{equation} \label{defc}
C(x-y)=  \left< (a^{\dagger}_x a_x + b^{\dagger}_x b_x)(a^{\dagger}_y 
a_y + b^{\dagger}_yb_y)\right>
\hspace{12pt}.
\end{equation}
The probability that a site is singly occupied is
\begin{equation}\label{sone}
S_1=\frac{\chi (0)}{\beta}
\hspace{12pt}.
\end{equation}
Furthermore
\begin{equation}\label{pp}
P(0)=1-S_1
\hspace{12pt}
\end{equation}
is the probability of zero or double occupancy.

The correlation function in Fourier space of the
quantity $E$ denoted by $\widetilde{E}$ is defined as
\begin{equation} \label{etilde}
\widetilde{E}(q)= \sum_z e^{iq z} E(z)
\hspace{12pt}.
\end{equation}

An indicator of the diverging correlation length at the
phase transition, which has been used e.g. in \cite{Scal,Morscal} is
\begin{equation} \label{ptilde}
\widetilde{P}_0  \equiv \widetilde{P}\left(q=(0,0)\right)
\hspace{12pt}.
\end{equation}
In the same way, the charge density wave indicator is $\widetilde{C}(\pi,\pi)$
and the uniform spin susceptibility is given by $\widetilde{\chi}(0,0)$.

From the $SU(2)^{\prime}$ symmetry, it follows that at $\mu=0$ the
Hamiltonian is invariant in particular under the transformation
\begin{equation}
(a_x,b_x) \to (\alpha_x,\beta_x)
\end{equation}
with
\begin{equation}
a_x = \frac{1}{\sqrt{2}} \left( \alpha_x + \beta_x^{\dagger} (-)^{x} \right) 
\hspace{12pt},
\end{equation}
\begin{equation}
b_x = \frac{1}{\sqrt{2}} \left( \beta_x - \alpha_x^{\dagger} (-)^{x} \right) 
\hspace{12pt}.
\end{equation}

Using that $\mu = 0$ implies half filling, i.e. $\left< n \right> = 1$,
this leads to
\begin{equation}
P(z) = (-)^z \left[ C(z) - 1 \right]
\hspace{12pt}.
\end{equation}

The model is soluble in the two limiting cases
$U=0$ (the free case) and $k=0$ (the atomic limit).
For reference, these
solutions and the corresponding values of the correlation functions
are given in Appendix A.
%
%
\section*{3. The Path Integral Formulation}
%
%
For the numerical simulation we use the path integral
formalism, as de\-scribed in Ref. \cite{MC}.
We begin with the Trotter break-up
\begin{equation}
e^{-\beta H}= \left(e^{-\frac{\beta H}{N_t}} \right)^{N_t} 
\end{equation}
and label each of the $N_t$ factors by an imaginary time index $t$
running from 1 to $N_t$.
One can then split the hopping term from the interaction term
\begin{eqnarray}
e^{-\frac{\beta H}{N_t}} & = & 
\exp \left[ \frac{k \beta}{N_t} \sum_{\left< x,y \right>}
\left( a^{\dagger}_x a_y +  a^{\dagger}_y a_x +
 b^{\dagger}_x b_y +  b^{\dagger}_y b_x \right)
\right] \nonumber \\
&& \times
\exp \left[
 \frac{U \beta}{N_t} \sum_x \left( a^{\dagger}_x a_x
             -\frac{1}{2} \right)
             \left( b^{\dagger}_x b_x 
               -\frac{1}{2} \right)
     + \frac{\mu \beta}{N_t} N  \right]
\hspace{6pt},
\end{eqnarray}
making an error of order $\frac{k  U \beta^2}{N_t^2}$. As a next step 
we introduce a scalar field $\left\{ \sigma_{x,t} \right\}$
for the time slice $t$ in order
to eliminate the four fermion term
\begin{eqnarray}  \label{aslice}
e^{-\frac{\beta H}{N_t}} & = &
\frac{1}{\left( {2\pi} \right)^{V/2}} \int 
 \prod_{x} \left[ \mbox{d} \sigma_{x,t} \right]
e^{-\frac{1}{2}\sum_x \sigma_{x,t}^2}
\exp \Bigg[
 \frac{k \beta}{N_t}     
\sum_{\left< x,y \right>}
\left( a^{\dagger}_x a_y +  a^{\dagger}_y a_x \right)
\nonumber  \\
&& + \frac{k \beta}{N_t}     
 \sum_{\left< x,y \right>}
  \left( b^{\dagger}_x b_y +  b^{\dagger}_y b_x \right)
\nonumber  \\
&& + \sum_x \left( 
a^{\dagger}_x a_x +  
b^{\dagger}_x b_x \right) 
\left\{ \sqrt{ \frac{U \beta}{N_t} } \sigma_{x,t} +
\left( \mu - U \right) \frac{\beta}{N_t} \right\} \Bigg] 
\hspace{12pt}.
\end{eqnarray}

Furthermore, we use the two relations
\begin{equation}
\exp \left[ a_x^{\dagger} a_y \right] = 1 + a_x^{\dagger} a_y
, x \neq y  
\; \; \mbox{  and  } \;\; 
\exp \left[ \alpha a_x^{\dagger} a_x \right] 
= 1 + a_x^{\dagger} a_x 
     \left[ e^{\alpha} - 1 \right] \hspace{12pt}, 
\end{equation}
and the same relations for $b^{\dagger}$ and $b$
to transform the sums in the exponential of Eq. (\ref{aslice}) 
\begin{equation}
\exp \left[ \frac{k \beta}{N_t} \sum_{\left< x,y \right> }  
\left( a^{\dagger}_x a_y \right) \right] =
\prod_{\left< x,y \right> } \left( 1 + \frac{k \beta}{N_t}    
a^{\dagger}_x a_y \right) + O \left( \frac{k \beta}{N_t}\right)^2
\end{equation}
and
\begin{eqnarray}
\exp \left[ \sum_x a^{\dagger}_x a_x \left( 
\sqrt{ \frac{U \beta}{N_t} } \sigma_{x,t} + 
{\rm const.}\right) \right]   =  \qquad\qquad\qquad & & \nonumber \\
\prod_{x} \left( 1 + a^{\dagger}_x a_x
\left( \exp \left[ \sqrt{ \frac{U \beta}{N_t} } \sigma_{x,t} +
{\rm const.} \right] -1 \right) \right)  & &
\hspace{12pt}. 
\end{eqnarray}

For every time slice we put the appropriate factors together 
and use the anticommutation relations 
to transform the expression into a sum of normal ordered operators,
neglecting terms
of higher order in $1/N_t$. The trace of the product
of normal ordered expressions 
in the fermion operators can be transformed into Grassmann
integrals in the standard fashion \cite{MC}, using
\begin{equation} \label{tr1}
\mbox{Tr} \left[ \prod_{i=1}^{N_t} 
: f_i \left( a^{ \dagger }, a \right): \right] =
\int  \prod_{t=1}^{N_t}
\left[ \left( \mbox{d} \bar{\eta}_t \right) f_t
\left( \bar{\eta}_t ,\eta_t \right)
\left( \mbox{d} \eta_t \right)
e^{ \bar{\eta}_t \left( \eta_t - \eta_{t-1} \right)} \right]
\end{equation}
with the boundary condition $\eta_0 = - \eta_{N_t}$
and similarly for $f_i (b^{\dagger}, b)$.  Finally, we
obtain the partition function $Z$, up to an unimportant numerical 
coefficient, as
\begin{eqnarray} \label{defz}
Z &= & \int   \prod_{x,t}  \left[ \mbox{d} \sigma_{x,t} \right] 
e^{-\frac{1}{2} \sum_{x,t}\sigma^2_{x,t}}
 \int \prod_{x,t} \left[ \mbox{d} \eta_{x,t} \mbox{d} \bar{\eta}_{x,t} \right]
   \prod_{x,t} \left[ \mbox{d} \theta_{x,t} \mbox{d} \bar{\theta}_{x,t} \right]
\nonumber \\
 && \ \exp \Bigg[
 \frac{k \beta}{N_t} \sum_{ \left< x,y \right>,t} \left( \bar{ 
\eta}_{x,t}  \eta_{y,t} + \bar{ \eta}_{y,t}  \eta_{x,t} \right) \nonumber \\
&& \qquad  + \sum_{x,t} \bar{ \eta}_{x,t} \left( \eta_{x,t} -\eta_{x,t-1}
 \right) \nonumber \\
&& \qquad  + \sum_{x,t} \bar{ \eta}_{x,t}  \eta_{x,t} \left( \mbox{exp}
 \left[ \sqrt{ \frac{U \beta}{N_t} } \sigma_{x,t} 
-(U- \mu) \frac{\beta}{N_t} \right] - 1\right) 
\nonumber \\
&+ & \ \ \left\{ \eta \to \theta \right\}  \Bigg]
\hspace{12pt},
\end{eqnarray}
when the $\bar{\theta}, \theta$ Grassmann variables
are associated with $b^{\dagger},b$.
This defines a fermionic matrix $M$, specified
for each set of $\sigma$-fields
by its elements between the
vectors $\bar{\eta}_{y,t^{\prime}}$ and $\eta_{x,t}$
\begin{eqnarray} \label{matrix1}
 \bar{\eta}M\eta &=& \frac{k \beta}{N_t} \sum_{ \left< x,y \right>,t}
 \left( \bar{ \eta}_{x,t}  \eta_{y,t} + \bar{ \eta}_{y,t}  \eta_{x,t} \right)
- \sum_{x,t} \bar{ \eta}_{x,t} \eta_{x,t-1}  
\nonumber \\
&& + \sum_{x,t} \bar{ \eta}_{x,t}  \eta_{x,t}  \mbox{exp}
 \left[ \sqrt{ \frac{U \beta}{N_t} } \sigma_{x,t} 
-(U- \mu) \frac{\beta}{N_t} \right]  
\hspace{12pt}.
\end{eqnarray}

The integrals over the Grassmann variables may be performed,
and we get
\begin{equation} \label{zdet}
Z =
    \int  \prod_{x,t}  \left[ \mbox{d} \sigma_{x,t} \right] 
e^{ - \frac{1}{2} \sum_{x,t}\sigma^2_{x,t}}
\left(\mbox{det}M\right)^2
\hspace{12pt}.
\end{equation}
Furthermore $\mbox{det}M=\left(\mbox{det} M\right)^{\dagger}$, so that
$\left(\mbox{det} M \right)^2= \mbox{det}\left(M^{\dagger}M \right)$,
which is useful for the construction of the algorithm.
It is possible to include the terms of order 
$1/N^2_t$ in the action. 
They are given in Appendix B and included in the simulation.

In order to compute expectation values of observables, it is convenient to
first consider that of the operator
\begin{equation}
O \equiv \exp \left[ - a^{\dagger} j \right] \exp \left[ \bar{j} a \right]
\end{equation}
when $\bar{j},j$ are Grassmann sources for the operators $a, a^{\dagger}$
taken at equal time, say $t=0$. Inserting $O$ as an additional factor into
the l.h.s. of Eq. (\ref{tr1}) corresponds to insert the factor $\widetilde{O}$
into its r.h.s., where
\begin{eqnarray}
\widetilde{O} & = &  \int \mbox{d} \eta_{N_t+1} \mbox{d} \bar{\eta}_{N_t+1}
   \exp \left[ \bar{j} \eta_{N_t+1} + \bar{\eta}_{N_t+1}j \right] \nonumber \\
 & \times &  \exp \left[ \bar{\eta}_{N_t+1} \left( \eta_{N_t+1} - \eta_{N_t}
              \right) \right] \nonumber \\
 & \times & \exp \left[ \bar{\eta}_{1} \left(  \eta_{N_t+1} - \eta_{N_t}
              \right) \right]
             \hspace{12pt}.
\end{eqnarray}   

This is obtained by associating to $\left( a, a^{\dagger} \right)$ an additional
Grassmann pair $\left( \eta_{ N_t + 1}, \bar{\eta}_{N_t + 1} \right)$, taking
care of the boundary conditions which now says $\eta_0 = - \eta_{N_t + 1}$
instead of $\eta_0 = - \eta_{N_t}$ (this explains the last exponential above).
Explicit integration yields
\begin{equation}
\widetilde{O} = 
 \exp \left[ \bar{j} \eta_{N_t} + \bar{\eta}_{1} j + \bar{j}j \right]
\hspace{12pt}.
\end{equation}
In fact $O$ could have been inserted in the trace between any two successive
time slices $t$ and $t+1$ (with the same result for the trace due to cyclic
invariance). The resulting $\widetilde{O}$ would then read
\begin{equation}
\widetilde{O} =
 \exp \left[ \bar{j} \eta_{t} - \bar{\eta}_{t+1} j + \bar{j}j \right]
\hspace{12pt}.
\end{equation}

This technique generalizes to the computation of any (time dependent)
correlator $\zeta$. After $\zeta$ is rewritten in normal order (any creation
operator brought to the left of any annihilation operator  by repeated use
of their commutation relations), one makes the replacement in
$\zeta \left( \left\{ a^{\dagger} \right\} \left\{ a \right\} \right)$
\begin{equation}
\left\{ a_x^{\dagger} \left[ t \right]  ,  a_y \left[ u \right] \right\}
\to
\left\{ \frac{\partial}{\partial j^{\left( a \right)}_{x,t+1}} ,
       \frac{\partial}{\partial \bar{j}^{\left( a \right)}_{y,u}} \right\}
\hspace{12pt},
\end{equation}
the similar one for the $b$'s, and ends up with the final result:
\begin{equation}
\left< \zeta \right> =
          \zeta \left(  \left\{ \frac{\partial}{\partial j} \right\}, 
                   \left\{ \frac{\partial}{\partial \bar{j}} \right\} \right)
                    \mbox{ln} Z \left( \bar{j},j \right)  \Bigg|_{j=\bar{j}=0}
\hspace{12pt},
\end{equation} 
where
\begin{eqnarray}
Z\left(\bar{j},j\right) = \int && \prod\limits_{x,t}
 \left[ \mbox{d} \sigma_{x,t} \exp 
\left( - \frac{\sigma_{x,t}^2}{2}\right) \right]
\times \left( \mbox{det}M \right)^2 \times \nonumber \\
\times && \exp \left[  \bar{j}^{\left( a \right)} M^{-1} j^{\left( a \right)}
                + \sum\limits_{x,t} \bar{j}_{x,t}^{\left( a \right)}
                     j_{x,t+1}^{\left( a \right)} + a \leftrightarrow b \right]
\hspace{12pt}.
\end{eqnarray}

For the observables of interest, listed 
in Eqs. (\ref{defn}, \ref{defp} - \ref{defc}),
the following expressions then follow.

\begin{equation} \label{defnup}
\left< n \right>  =  2 - 2  \left< M^{-1}_{x,N_t; x,1} \right>
\hspace{12pt},
\end{equation}

\begin{equation} \label{pmat}
P(x-y) = \delta_{xy} \left[ 1 - 2 \left< M^{-1}_{x,N_t; x,1} \right> \right]
  + 2 \left< \left( M^{-1}_{x,N_t; y,1} \right)^2 \right>
\hspace{12pt},
\end{equation}

\begin{equation}  
{\chi(x-y)\over\beta} = 2  \delta_{xy} \left< M^{-1}_{x,N_t; x,1} \right>
       -2   \left< M^{-1}_{x,N_t; y,1}   M^{-1}_{y,N_t; x,1} \right>
\hspace{12pt},
\end{equation}

\begin{eqnarray}  \label{defC}
C(x-y) = && 2 \delta_{xy}  \left< M^{-1}_{x,N_t; x,1} \right> 
         + 4 - 8  \left< M^{-1}_{x,N_t; x,1} \right> \nonumber \\
     &+ & 
     4  \left< M^{-1}_{x,N_t; x,1}   M^{-1}_{y,N_t; y,1} \right>
         -2   \left< M^{-1}_{x,N_t; y,1}   M^{-1}_{y,N_t; x,1} \right>
\hspace{12pt}.
\end{eqnarray}

We will also consider the average value of the field $\sigma_x^{\prime}$
defined as 
\begin{equation} \label{sigpri}
\sigma_{x,t}^{\prime} = \sqrt{\frac{N_t}{\beta}} \sigma_{x,t}
\hspace{12pt}.
\end{equation}
Indeed, from the equation of motion for the $\sigma$-fields, 
which can easily be derived from Eq. (\ref{aslice}) or  (\ref{defz}), we have
\begin{equation} \label{siginf}
\lim\limits_{N_t \to \infty} \left< \sigma_{x,t}^{\prime} \right>
= \sqrt{U} \left< n \right>
\hspace{12pt}.
\end{equation}
This relation, together with $\left< n \right> =1 $ at $\mu = 0$ in the
large $N_t$ limit (at finite $N_t$
the particle-hole symmetry is broken 
\cite{MC}), will be used as a check that we actually
considered large enough $N_t$ values. 

%
%
\section*{4. The numerical simulation}
%
%
The expectation values of (\ref{defnup}-\ref{sigpri}) are obtained from
$\sigma$-field
configurations corresponding to the partition function (\ref{zdet}).
These configurations are generated with an Hybrid Monte Carlo algorithm
(HMC-algorithm) which is a widely used tool to simulate systems involving
the fermion matrix (e.g.  \cite{MC}, \cite{GKS}, \cite{GIKP}).
Its main virtues are that it is an exact algorithm and that, at least in
principle, the computer time grows slowly with system size.

Introducing one set of pseudofermion fields $\phi^{(i)}$ for each species
$\eta$ and $\theta$
and momenta $p$ conjugate to the $\sigma$ fields,
the  Hamiltonian of the system can be written as
\begin{equation}\label{hfic}
{\cal H} = \frac{1}{2} \sum\limits_j p_j ^2 +
 \frac{1}{2}   \sum\limits_j \sigma_j^2 + 
\sum_{i=1}^2   \sum\limits_{j,k}  {\phi_j^{(i)}} 
\left( M^{\dagger} M \right)_{jk}^{-1} \phi_k^{(i)}
\hspace{12pt},
\end{equation}
where $j$ and $k$ denote sites on the $N_s^2 \times N_t$ lattice.
Starting from a configuration $\sigma$, a new configuration is obtained
in three steps.
\begin{itemize}
\item[i)] First the pseudofermion fields are generated from gaussian distributed
vectors $r^{(i)}$ by $\phi^{(i)}= M^{\dagger}(\sigma ) r^{(i)}$
 and gaussian momenta $p$ are chosen.
\noindent
\item[ii)] Then $\sigma$ and $p$ are updated by molecular dynamics with
${\cal H}$ as Hamiltonian, the set of Hamilton's equations being solved with a
discrete time step $\mbox{d} \tau$. So $\cal{H}$ is a constant of motion
in the limit $\mbox{d} \tau \to 0$. 
The integration of the molecular dynamics equation
is performed by a leapfrog method \cite{GKS}
\begin{eqnarray} \label{leapfrogs}
\sigma_j |_{ (n+1){\rm d} \tau} &=& \sigma_j |_{n {\rm d} \tau} +
 \mbox{d} \tau \ p_j - \frac{(\mbox{d} \tau)^2}{2}
 \left. \frac{\partial {\cal H}}{\partial \sigma_j}\right|_{n {\rm d} \tau}
 \hspace{12pt}, \\
p_j|_{ (n+1){\rm d} \tau} &=& p_j|_{ n{\rm d} \tau}
   -\frac{ \mbox{d} \tau}{2} \left(
\left. \frac{\partial {\cal H}}{\partial \sigma_j} \right|_{n{\rm d} \tau}+
 \left. \frac{\partial {\cal H}}{\partial \sigma_j}\right|_{ (n+1){\rm d} \tau}
\right) 
\hspace{12pt}.\label{leapfrogp}
\end{eqnarray}
\noindent
\item[iii)] After $n_{\rm MD}$ steps, the resulting $\sigma$ configuration
is accepted or refused by a Metropolis test.
\end{itemize}

This algorithm introduces two parameters, the number of steps $n_{\rm MD}$ and
the step size ${\rm d}\tau$. The correlation between configurations decreases
when the trajectory length $\tau = n_{\rm MD}\mbox{d} \tau$ increases, while
increasing $n_{\rm MD}$ increases the computation time and increasing
$\mbox{d} \tau$ decreases the acceptance. We have taken $\tau\sim 1.0$, (more
precisely between 0.8 and 1.6) with $n_{\rm MD}$ such that a large enough
acceptance is reached for all the volumes used (the acceptance decreases with
increasing volume \cite{MC},\cite{GKS}).
This is obtained with $n_{\rm MD}\sim 4\beta U$.
We also tested trajectories with Poisson-distributed lengths,
i.e. $n_{\rm MD}$ distributed around a certain mean value and
$\mbox{d} \tau$  held fixed \cite{Mack,Wein}.
This does not decorrelate measurements of
observables compared to $\tau$ fixed.
The only effect is to increase the acceptance a few percent if
$n_{\rm MD}$ is small.

For an observable $O$  whose value at Monte Carlo time is $O_i$ the
autocorrelation function is
\begin{equation}
C_O(t) = \frac{\left< O_i O_{i+t} \right>- \left< O_i\right>^2 }
{\left< O_i^2\right> -\left< O_i\right>^2 }
\hspace{12pt}.
\end{equation}

We estimated the autocorrelation time $\xi_O$ by
\begin{equation}
\xi_O =\frac{1}{t_c}\sum_0^{t_c-1}\ln\left[\frac{C_O(t+1)}{C_O(t)}\right]
\hspace{12pt},
\end{equation}
disregarding the noisy region $t>t_c$ where $C_O(t)$ is less than 0.05.
This gives the exact autocorrelation time if $C_O$ is purely exponential.
The integrated autocorrelation time
\begin{equation} \label{tint}
t_{\rm int}=\sum_0^{t_c-1}C_O(t)
\hspace{12pt},
\end{equation}
coincides with $\xi_O$ within 5\% under the same assumption.

The physical and algorithmic parameters are given in
Tables (\ref{algo4},\ref{algo28}) along with the number of trajectories
simulated, the autocorrelation time $\xi_{\sigma^{\prime}}$
and the acceptance.
Other autocorrelation times connected to fermionic quantities are always
very small as compared to $\xi_{\sigma^{\prime}}$.
With the parameters chosen, the smallest values of the acceptance
at $U=4$ are
0.64 for the lattice $8^2\times 64$ at $\beta =6$ and 0.71 for the
lattice $8^2\times 32$ at $\beta =4$.

The updates of $\sigma_j$ and $p_j$ (Eqs. \ref{leapfrogs}, \ref{leapfrogp})
require twice the computation of a vector
$u = \left( M^{\dagger} M \right)^{-1} \phi$
for a given $\phi$. This computation, which costs most of the computer time,
is performed by a conjugate gradient subroutine with a diagonal
preconditioning \cite{GL}, which speeds up the convergence considerably.
The conjugate gradient iteration is stopped after $c_{cg}$ steps,
when $\left|  \left( M^{\dagger} M \right) u_{c_{cg}} - \phi \right|^2$
becomes less than $10^{-10}$ times the value obtained for the
initial guess  $u_0$.

The number of conjugate-gradient iterations can be significantly reduced if
we use for $u_0$ the value obtained at the previous molecular dynamics step.
However this method leads to dramatic decreases of the acceptance when
increasing $\beta$ and $V$. So we always start with $u_0 = 0$, in
such a way that no peculiar direction of phase space is favored or suppressed by
the numerical uncertainties.

The number of conjugate gradient iterations needed to reach the prescribed
accuracy ranges between $\sim 30 \%$ and $\sim 65 \%$ of the vector length.
However in some cases, this accuracy is not obtained in $N_s^2 \times N_t$
iterations, in particular for small volumes. Such situations can be avoided
by reducing the trajectory length $\tau$, and this explains the values
$\tau = 0.8$ and 1.2 in Table (\ref{algo4}). However, this comes with the
price of increasing the $\sigma$ autocorrelation times.

The matrix $M$ used in this computation is given in Eq. (\ref{matrix2}) of
Appendix B.  It is valid to order $1/N_t^2$. Some of the effects of using
this second order matrix instead of the first order matrix of
Eq.~(\ref{matrix1})  are shown in Table (\ref{compordre}).
At $\mu = 0$, and for $N_t$ large enough, one expects $\left< n \right>$ to
approach 1 and, according to Eqs. (\ref{sigpri},\ref{siginf}), 
$\left< \sigma^{\prime}_{x,t} \right>$ to approach $\sqrt{U}$. This is 
clearly better achieved with the improved action (\ref{matrix2}).
While $\widetilde{P}_0$ is not affected within errors by the change of
the action, $\widetilde{\chi}$ unexpectedly happens to depend more on
$N_t$ in the second order case.

Because increasing $N_t$ is costly in computer time, we currently used
$N_t = 2 \beta U$. The exceptions, apart for analysis of $N_t$ dependencies,
concern the case $\beta=0.5$, $U=8$ $(N_t=32)$ and the low temperature points
$\beta=6$, $U=4$ $(N_t=64)$. 

The results obtained for $\left< \sigma^{\prime} \right>$, filling of the
band $\left< n \right>/2 $, pairing correlation  $\widetilde{P}_{0}$,
spin susceptibility $\widetilde{\chi}$ and single occupation probability
$S_1$ for various lattices and values of the inverse temperature $\beta$
and chemical potential $\mu$ are given in Tables (\ref{Res4},\ref{Res28}).
The errors reported therein are calculated by incorporating the
autocorrelation time $t_{\rm int}$ of Eq. (\ref{tint}) by using a variance
which is the naive variance multiplied by $(1 + 2 t_{\rm int})$.

We cannot compare the efficiency of our algorithm with 
Quantum Monte Carlo \cite{WS}, because for that algorithm
we do not have the information
needed.

%
%
\section*{5. The S-Wave Pair-Field Correlation Function}
%
%
One of the
observables of interest in the context of a superconducting phase transition
is the pairing correlation function $P$ and its
Fourier transform  $\widetilde{P}$.  They are defined
by (\ref{defp}) and (\ref{etilde}) respectively.
In the present work we have
studied the 
$q=(0,0)$ equal-time s-wave pair-field correlation
function $\widetilde{P}_0$, Eq. (\ref{ptilde}), which has previously been
examined e.g. in \cite{Scal,Morscal}.
For this quantity we obtain from Eq.(\ref{pmat})
\begin{equation} \label{psim}
\widetilde{P}_0 = 1 - 2 \left< M^{-1}_{x,N_t;x,1} \right> 
+ 2 \sum\limits_y 
\left< \left( M^{-1}_{y,N_t;x,1} \right)^2 \right> 
\hspace{12pt}.
\end{equation}

Let us recall \cite{Scal,Morscal} what can be expected
on general grounds for the behavior of $\widetilde{P}_0$ as a function
of $\beta$, $U$, $\mu$ and $N_s$, the linear size of the lattice.

A first remark is that under a global particle-hole transformation
\begin{equation}
a_x \to (-)^x a_x^{\dagger}
\end{equation}
\begin{equation}
b_x \to (-)^x b_x^{\dagger}
\end{equation}
which changes $H(\mu)$ of Eq. (\ref{hubham}) into $H(-\mu)$, up to an
irrelevant additive constant, the operator defining $P$
in Eq. (\ref{defp}) remains
unchanged. Hence
\begin{equation}
\widetilde{P}_0(\mu) =  \widetilde{P}_0(- \mu)
\hspace{12pt},
\end{equation}
whereas the filling density verifies
\begin{equation}
\left< n \right> (\mu) = 1 - \left< n \right> (-\mu)
\hspace{12pt}.
\end{equation}

In the vicinity of $\mu = 0$, and above any transition temperature which could
lead to non-analyticities, we thus expect $\widetilde{P}_0$ to be a function 
of $\mu^2$, while $\left< n \right> (\mu)$ is linear in $\mu$.

At $\mu=0$, because of the $SU(2)^{\prime}$ symmetry (Eq. \ref{sup}), no
transition may occur at $T \neq 0$, and as $T \to 0$, the correlation length
$\xi$ and $\widetilde{P}_0$ are expected to diverge according to
\begin{equation} \label{H1}
\xi \propto \exp \left( A /T \right) 
\end{equation}
\begin{equation}  \label{H2}
\widetilde{P}_0 \simeq b \xi^2
\hspace{12pt}.
\end{equation}

At $\mu \neq 0$, the  $SU(2)^{\prime}$ symmetry  is broken down to a $U(1)$
symmetry associated with the particle number, which can lead to a finite $T$
transition of the Kosterlitz-Thouless \cite{KT} type. Then
the expectation is, above the critical temperature $T_{\rm KT}$,
\begin{equation} \label{KT1}
\xi_{\rm KT} \propto \exp \left[ A /\sqrt{ T - T_{\rm KT} (\mu,U) }\right] 
\end{equation}
and
\begin{equation}  \label{KT2}
\widetilde{P}_0 \simeq b \xi_{\rm KT}^{2 - \eta_{\rm KT}}
\hspace{12pt},
\end{equation}
with $\eta_{\rm KT}=.25$. Of course $A$ and $b$ are functions of $\mu$ and $U$.
Also $T$
dependent prefactors may occur in these expressions.
But it is reasonable to think that the $\mu$,$U$ dependences
of  $\xi_{\rm KT}$ and $\widetilde{P}_0$ are 
dominated by that of $T - T_{\rm KT}$.

In a finite system of size $N_s^2$, standard scaling arguments predict
a behavior 
\begin{equation}  \label{HL}
\widetilde{P}_{0,N_s} \simeq N_s^2 f(N_s/\xi) \mbox{ at } \mu = 0 
\hspace{12pt},
\end{equation}
and
\begin{equation}  \label{KTL}
\widetilde{P}_{0,N_s} \simeq N_s^{2-\eta_{\rm KT}}
  f(N_s/\xi_{\rm KT}) \mbox{ at } \mu \neq 0 
\end{equation}
for the pair correlation, in a domain of parameters $\beta, \mu, U$ where
the $\xi$'s are large, and $N_s/\xi$ is kept fixed. 

On the basis of numerical QMC simulations of the model, Ref. \cite{Morscal}  
confirmed this overall picture and concluded that a maximum transition
temperature of order 0.2 is reached for $U \simeq 8$ and
$\left< n \right> \simeq .85$, while at $U=4 $ the critical temperature
$T_{\rm KT}$ is of order 0.1 and 0.05 respectively
at $\left< n \right>$ = 0.87 and 0.5.

We will now discuss the results of our numerical HMC simulations.
We start with our results at $U=4$, where we have most of the data.

We first focus on the $\mu$-dependence of $\widetilde{P}_{0}$, more convenient
than its filling dependence because  $\left< n \right> (\mu)$
suffers from finite size effects which are not of the same nature as those
we are interested in, being important even in the non interacting case for
the $N_s$ values commonly considered. At fixed $T$ and $U$, the above
discussion indicates that $\widetilde{P}_{0}$ is flat close to $\mu=0$,
and should increase with $| \mu | $ as a consequence of $T - T_{\rm KT}(\mu,U)$
decreasing, a maximum of $T_{\rm KT}$ leading to a maximum of
$\widetilde{P}_{0}$ in the same $\mu$ region. This qualitative behavior should
survive in a finite, large enough box provided the function $f(x)$ of
Eq. (\ref{KTL}), which is proportional to $x^{-1.75}$ at large $x$, stays
monotonic.

\begin{figure}[htbp]
\epsfig{file=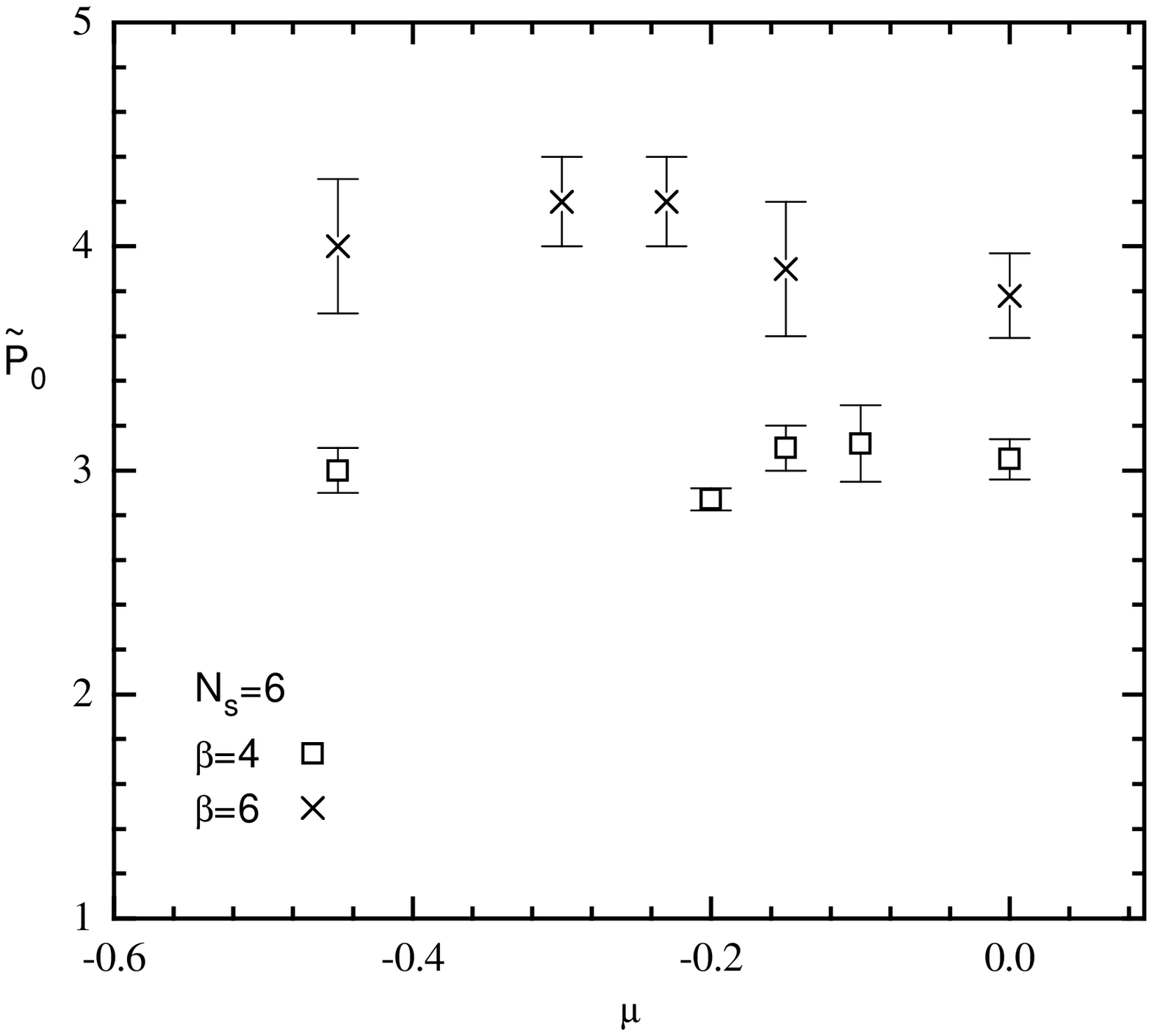, width=130mm}
\caption{ $\widetilde{P}_0$  as a function of $\mu$ for $U=4$.}
\label{p06}
\end{figure}

In order to probe such a behavior, we select out from our data of
Table (\ref{Res4}) the values of $\widetilde{P}_{0,N_s}$ obtained at
the largest $\beta$ (4 and 6) and $N_s=6$
available for a sufficient range of $\mu$ values. The result is shown in
Fig. (\ref{p06}). These data are compatible with $\widetilde{P}_{0,N_s}$
being constant with respect to $\mu$, and not
compatible with a sharp maximum in this $\mu$ range (corresponding
to a filling between .65 and 1.0 for both $\beta$ values), as suggested in
Ref. \cite{Morscal}.

We may still examine the dependence of $\widetilde{P}_{0,N_s}$ on $N_s$ at
fixed $\mu$, to see if these data give evidence for a KT transition at
$\mu \neq 0$. We therefore fix $\mu$ at -0.15, a
value at which we have data in the range $1 \leq \beta \leq 6$ and
for various $N_s$ values in the 4-8 range.

\begin{figure}[htbp]
\epsfig{file=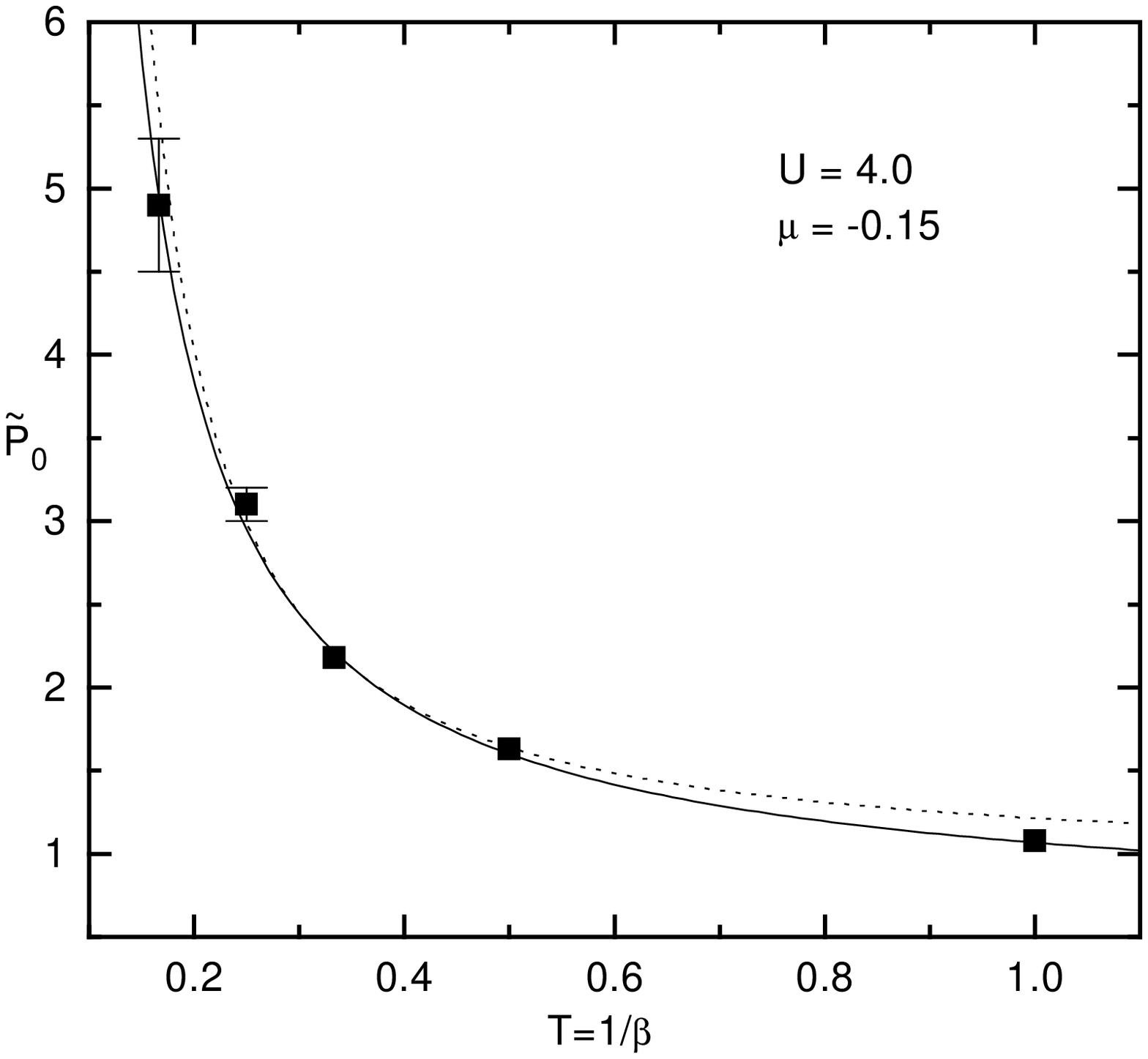, width=130mm}
\caption{ $\widetilde{P}_0$  as a function of the temperature for $U=4$ and
$\mu =-0.15$. The solid curve is the fit to
$\widetilde{P}_0 = b \exp \left[ a / (T - T_{\rm KT})^{1/2} \right]$,
the dashed curve is the fit to $\widetilde{P}_0 = b \exp \left[ a /T \right]$.
The parameters are given in line 1 and 4 of Table \ref{fit}.}
\label{fig3}
\end{figure}

Let us first {\it assume } that at each
$\beta$ the largest $N_s$ value available is large enough for the system 
to be close to its thermodynamical limit, an assumption which is
verified at $\beta=4$ as seen from our data at $N_s = 4,6,8$. In 
Fig. (\ref{fig3}) we show $\widetilde{P}_{0,N_s}$ as a function of $T$
for $N_s=4$ at $\beta \leq 3$ and $N_s=8$ for $\beta \geq 4$, together
with a fit to Eqs. (\ref{KT1}, \ref{KT2}) (continuous line). The
fit is remarkably good in the entire temperature range. It favors
a KT transition at $T_{\rm KT} = .03(2)$, at about 4 standard deviations
away from the value .1 preferred in Ref. \cite{Morscal}. Also shown
is a fit to Eqs. (\ref{H1}, \ref{H2})
(dotted line), which is compatible as well
with the data, provided the $T=1$ point is ignored.
Imposing $T_{\rm KT}=0.1$ gives poor fits (see Table (\ref{fit})).

\begin{figure}[htbp]
\epsfig{file=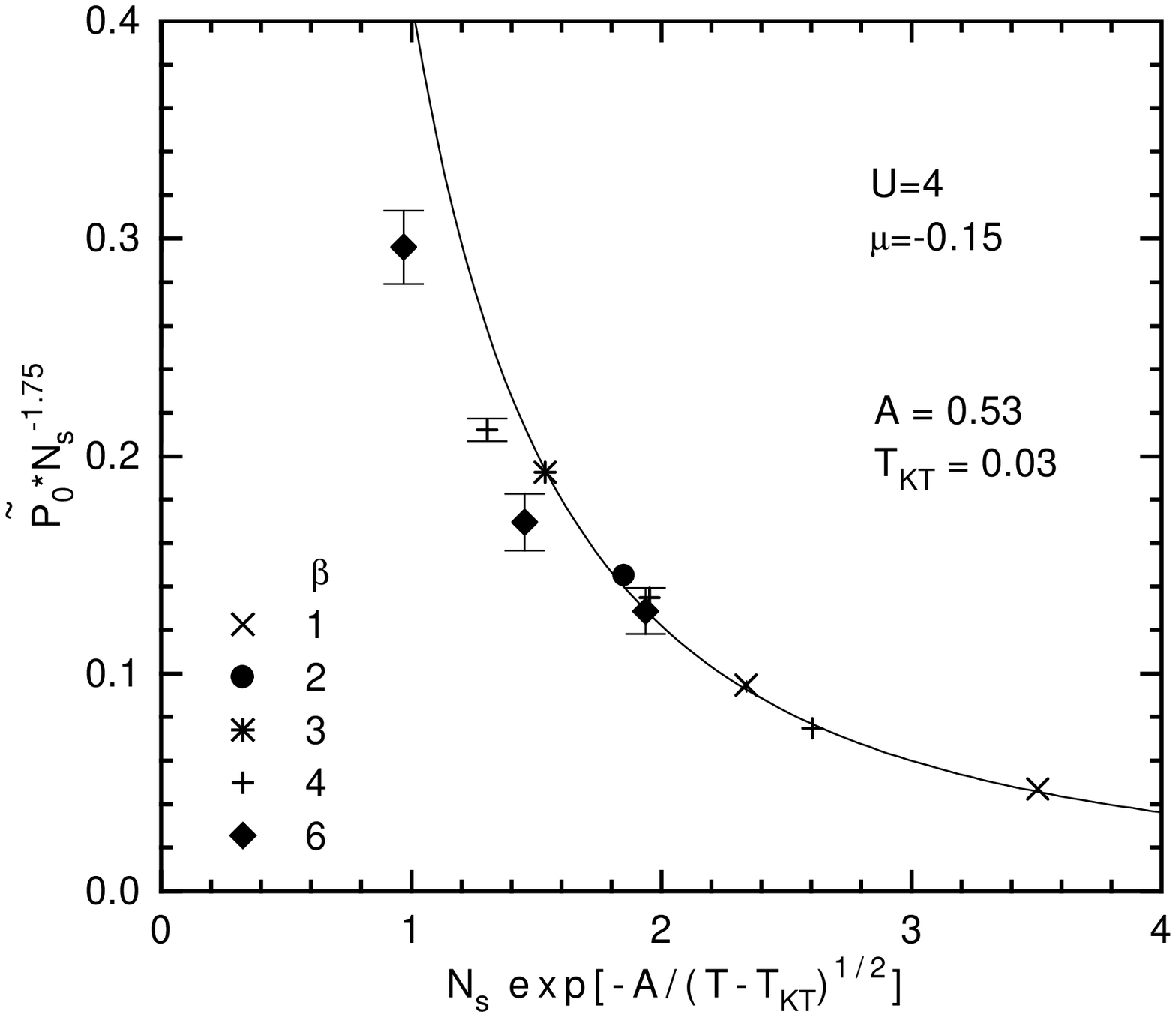, width=130mm}
\caption{Scaling of $\widetilde{P}_0$ (Eq. (\ref{KTL}) ).}
\label{pla}
\end{figure}

\begin{figure}[htbp]
\epsfig{file=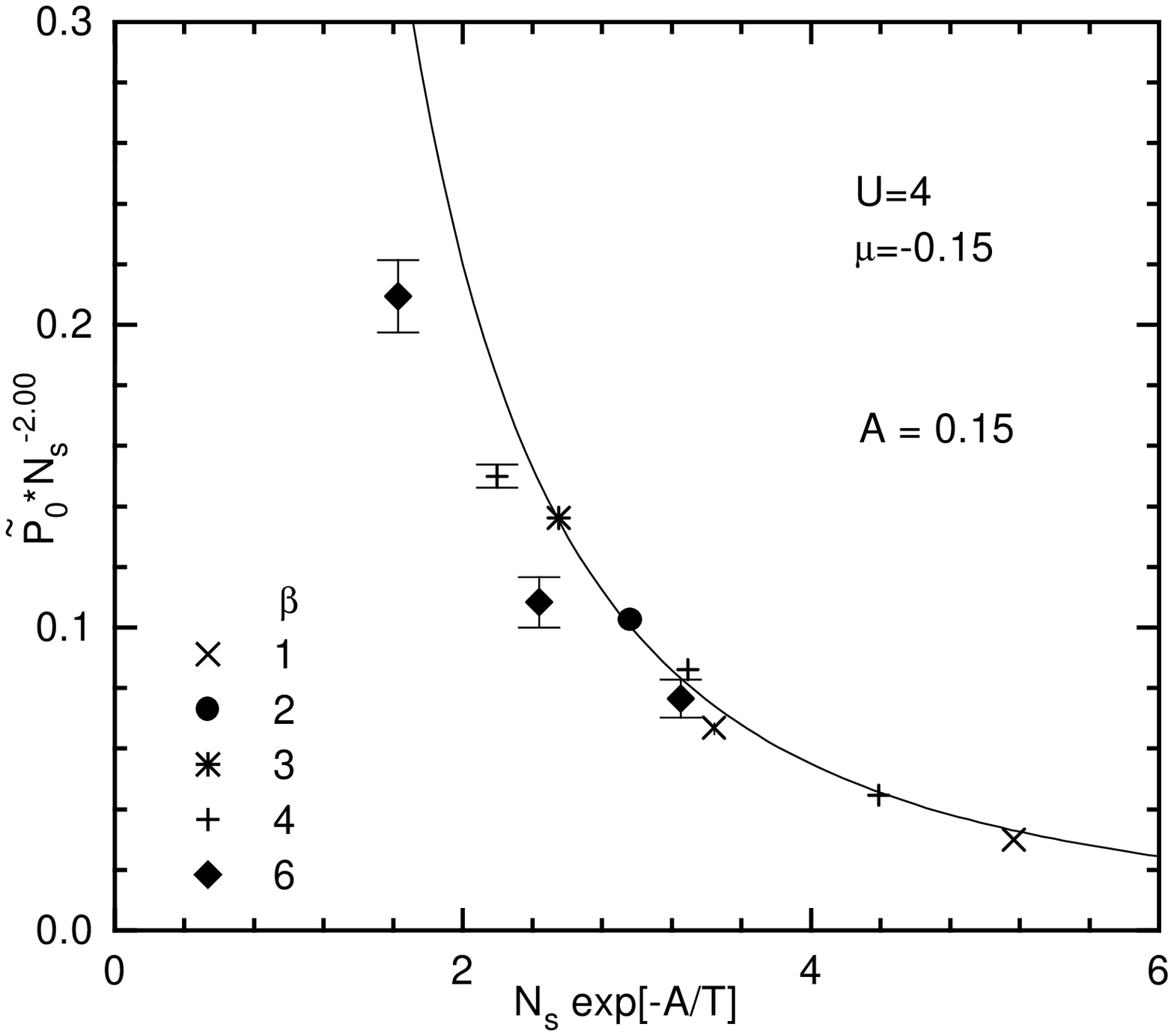, width=130mm}
\caption{Scaling of $\widetilde{P}_0$ (Eq. (\ref{HL}) ).}
\label{plb}
\end{figure}

These fits provide us with a guess for the parameterization of the
correlation length, up to a normalization factor, which allows us 
to test scaling behavior given by (\ref{KTL}) or (\ref{HL}). With the
parameters of lines 1 and 4 of Table (\ref{fit}), we obtain the results
shown in Figs. (\ref{pla}) and (\ref{plb}) respectively, together with the
asymptotic limits $c (N_s/\xi_{\rm KT})^{-1.75}$ and $c^{\prime}(N_s/\xi)^{-2}$
expected in the infinite volume limit for the scaling functions
$f_{\rm KT}$ and $f$ ($c$ and $c^{\prime}$ are determined by the 
$\beta = 3$, $N_s=4$ point). Since both figures may be considered
as suggestive
of a scaling behavior, the evidence for a Kosterlitz-Thouless transition
at $\mu = -0.15$ remains poor.

From this analysis, our conclusion is: If there is a Kosterlitz-Thouless
transition for $U=4$, $\mu = - 0.15$, it occurs at such a small value
of $T$ in units of the hopping parameter $k$ (a few percent) that its 
effects are hardly distinguishable from those of the $T=0$ singularity
expected at $\mu=0$. (Eq. \ref{H1}).

This conclusion is different from that of \cite{Morscal}, and one
must try to understand why. At the quantitative level, for $U=4$, the
values we find for $\widetilde{P}_{0,N_s}$ at $\mu=-0.15$ above $\beta=3$
are smaller than in  \cite{Morscal} at $\left< n \right> = 0.87$,
especially at $\beta=6$. That  $\left< n \right>$ depends on $N_s$ at
fixed $\mu$ as already mentioned (see Table (\ref{Res4})) is insufficient to
explain the difference since we do not observe significant $\mu$
dependence in $\widetilde{P}_{0,N_s}$.

Of course larger $\widetilde{P}_{0}$ values at a given $\beta$ favor larger
$T_{\rm KT}$ when analyzed in terms of the (\ref{KTL}) Ansatz, and this may
explain the value $T_{\rm KT} \sim .1$ proposed in \cite{Morscal}. However,
we would like to point out that the data corresponding to the highest $\beta$
values 8 and 10 of the latter reference were not used in the finite size
analysis provided there. A look at these data strongly suggests that including
them would have considerably lowered the guessed value of $T_{\rm KT}$.

We did not take enough data at $U=2$ and $U=8$ to allow for an analysis similar
to that achieved at $U=4$. A look at Table (\ref{Res28}) however confirms that
the pair correlation increases with $\beta$ at fixed $[ \mu, U ]$, while its
values at $\mu = 0$ and -0.15 are consistent with each other at fixed
$[ \beta, U ]$. Some information comes from comparing  $\widetilde{P}_{0}$
at $U=2,4$ and $8$ for $\beta=2$ and $\mu = 0.$ fixed. Its values are
respectively 1.127(4), 1.63(3) and 2.2(2). It would be incorrect to directly
infer from this that also the correlation length increases. In fact, as $U$
increases at fixed temperature, more pairs are formed and this affects the
normalization of the correlation function. As a measure of this normalization
we take the $x = y$ (zero distance) value of the correlation of
Eq. (\ref{defp}),
\begin{equation}
P(0) = 
\left<
a_x^{\dagger} b_x^{\dagger} b_x a_x + b_x a_x a_x^{\dagger} b_x^{\dagger}
\right>
\hspace{12pt},
\end{equation}
which is nothing but the average number densities of doubly
occupied and empty sites, related to the single occupation probability
$S_1$, given in Tables (\ref{Res4},\ref{Res28}), by $P(0) = 1 - S_1$.
Hence a better representative of the correlation length variations is  
$\widetilde{P}_{0}  / P(0)$, and for this quantity at $\beta=2, \mu=0,
U=2,4,$ and 8 respectively we find 1.80(2), 2.18(4) and 2.4(2). 

This leaves uncertain the guess that the correlation length is larger
at $U=8$ than it is at $U=4$.

The above remark about the normalization of $\widetilde{P}_{0}$, which might
be important for a discussion of the $U$ dependence of the KT transition
temperature, does not alter our previous analysis at $U=4$ fixed because,
as seen from Table (\ref{Res4}), $P(0)$ is essentially insensitive to $\beta$,
$\mu$ and $N_s$ in the explored ranges.  This is true even at temperatures as
high as 1 at $U=2$ and 4. The largest variation observed for $P(0)$ is found
between $T=0.5$ and $T=2$, for $U=8$, where it however does not exceed 10\%.
So, in the $U$ ( and $\mu$) range considered, the pairs are very
robust against temperature rises.

The independence of $S_1$ on the filling for fixed $U$ in the region
we investigated  means that the increase in the filling goes  into
the formation of pairs. This is, however, also true in the free case in this
region. This may be related to the fact that $S_1$, as well as
$\widetilde{P}_{0}$, is an even function of $\mu$.
\vfill\eject
%
%
\section*{6. The susceptibility: Comparison of numerical results with hopping
 parameter expansion}
%
%
The free energy of the repulsive Hubbard model has been expanded in
powers of the hopping parameter $k$ by various authors \cite{PW,DfB,OH,HOA}.
We start from the result of \cite{HOA}, where the expansion is pushed
to order 5 in $k^2$. Via a particle-hole transformation on one of
the two electron operators, the free energy expansion is transformed into
that relevant for the Hamiltonian in the attractive case Eq. (\ref{hubham}).
For arbitrary values of $\beta^{-1}$, $\mu$ and $h$ respectively of the
temperature, chemical potential and external magnetic field, we end up
with the following truncated expansion
( $F$ is the negative of $\beta$ times the free energy density) 
\begin{eqnarray} \label{Free}
F(\beta,U,\mu,h)&  \equiv  & \frac{1}{V} \mbox{ln} Z + O (k^{12})
 \nonumber  \\
& =  &  \mbox{log} z + 
\sum\limits_{\kappa = 1}^5  \sum\limits_{ \left\{ R \right\} }
\frac{v^{2 \kappa}}{\left( \beta U \right)^i } 
x^{i_\mu}    y^{i_h}  x^{i_w} C_{\kappa;  \left\{ R \right\} }
\hspace{12pt},
\end{eqnarray}
where
\begin{eqnarray} 
&&
x=e^{\beta \mu}  ,  y=e^{\beta h} ,   w = \exp \left[ - \frac{\beta U}{2}
 \right] \nonumber \\
&&
v = \frac{\beta k}{z} \nonumber \\
&&
z = 1 + x w \left( y + 1/y \right) + x^2 
\hspace{12pt}.
\end{eqnarray}
and $\left\{ R \right\}$ represents the set of integers
$\left\{ i, i_{\mu} ,i_{h} ,i_{w}\right\}$.

The symbol $C_{\kappa, \left\{R \right\}}$ represents the 3836 non zero 
coefficients provided in \cite{HOA}, labeled by the order
$\kappa$ in $k^2$ and the associated set $\left\{R \right\}$. 
By differentiation with respect to $\mu$ and $h$, one obtains
similar series (here taken at zero magnetic field) for the filling
density, the static uniform charge density correlation and the static
uniform spin susceptibility. We focus on the latter, and study the
truncated series
\begin{equation}
\frac{1}{\beta} \widetilde{\chi}(q=0) =
\left( y \frac{\mbox{d}}{\mbox{d}y} \right)^2 F \equiv
\sum\limits_{\kappa = 0}^{5} A_{\kappa} k^{2 \kappa}
\hspace{12pt}.
\end{equation}

The coefficients $A_{\kappa}$ follow from the expansion
(\ref{Free}) of $F$. Although it is not strictly speaking a high
temperature series, but a series in $(\beta k^2)$ with 
$\beta$-dependent coefficients, extracting information
from such short expansions at low temperature is notoriously
difficult. 

An idea of the problem at hand is given by the following examples.
We take $U=4$, $\mu=0$ and compute the coefficients $A_{\kappa}$
for $\beta=1$ and $2$. Their numerical ratios to $A_{0}$ are
shown in Table (\ref{x}).

One observes a quite discouraging pattern of orders of magnitude and
signs (remember we want to evaluate $\widetilde{\chi}$ at
$k = 1$). We tried to understand what happens by starting from
the free case $U=0$, at $\mu=0$.  The exact susceptibility
is given by (see appendix A)

\begin{equation} \label{chiex}
\widetilde{\chi} = \beta \int\limits_{0}^{\pi}
\frac{\mbox{d}p_1  \mbox{d}p_2 }{\pi^2}
\frac{1}{1 + \mbox{cosh} ( 2 k \beta \delta_p) } 
\end{equation}
with
\begin{equation}
\delta_p = -( \mbox{cos} p_1 + \mbox{cos} p_2  )
\hspace{12pt}.
\end{equation}
Its expansion in $(\beta k)^2$ to $5^{th}$ order is
found to be 
\begin{equation} \label{chi5}
\frac{2 \widetilde{\chi}}{\beta} =
1 - (\beta k)^2 + \frac{3}{2} ( \beta k)^4 
 - \frac{85}{36} ( \beta k)^6
 + \frac{1085}{288} ( \beta k)^8
 - \frac{4837}{800} ( \beta k)^{10}
\end{equation}

The signs now alternate, but again the size of the coefficients
in $k^2$ increases fast, the more so the temperature is lowered.
Here the origin of this blow up is clear from expression (\ref{chiex}).
Rewriting it as 
\begin{equation}
\frac{\widetilde{\chi}}{\beta} =
\int\limits_{0}^2 \mbox{d} \delta \rho (\delta)
\frac{1}{1 + \mbox{cosh} \left( 2 k \beta \delta \right) }
\end{equation}
where $\rho(\delta)$ represents the density of states,
and recalling that $\rho$ has only
a logarithmic (van Hove) singularity at $\delta=0$, all singularities
of  $\widetilde{\chi}$ in $k$ at finite distance are end point 
($\delta=2$) singularities. They are located at 
$\mbox{cosh} (4k\beta)=-1$, and the closest one, which fixes the radius 
of convergence, is at $( \beta k)^2 = - \pi^2 / 16$. Hence the
alternate signs in the above series and also the fact that
$\pi^2 / 16$ times the ratio $| A_{5} / A_{4}|$  is very
close to 1 (namely .990) . Finally, we note from the above integral that
the most singular part of $\widetilde{\chi}/ \beta$ is 
proportional to $\mbox{tanh}(2 \beta k)/(\beta k)$. This leads
us to perform the change of variable $\beta k \to u$ defined
by
\begin{equation} \label{defu}
u = \mbox{tanh} \left( 2 \beta k \right) / (\beta k) -2
\hspace{12pt},
\end{equation}
and to re-expand  $\widetilde{\chi}/ \beta$ in powers of $u$. We find
\begin{equation}
\frac{2 \widetilde{\chi}}{\beta} =
1 + \frac{ 3u}{8}    
- \frac{9 u^2}{640}
+ \frac{246 u^3}{51469}
- \frac{4096 u^4}{2238401}
+ \frac{30034 u^5}{40581943}
+ \cdots
\end{equation}
that is
\begin{equation} \label{chinum}
\frac{2 \widetilde{\chi}}{\beta} \simeq
1. + .375u -     1.41 \; 10^{-2} u^2
+ 4.78 \; 10^{-3} u^3
- 1.83 \; 10^{-3} u^4
+ 7.40 \; 10^{-4} u^5
 \end{equation}

The interval $\beta k \geq 0$ is mapped onto 
$-2 \leq u \leq 0$, and low temperatures mean $u \sim -2$. Clearly
this series in $u$ looks much more tractable than the original
one. Indeed, actual comparison of Pad\'e approximants of
(\ref{chinum}) with the exact result (\ref{chiex}) shows that, up to
$\beta \sim 10$, the latter can be approached with an accuracy
better than 5\%, an achievement not expected from the series (\ref{chi5})!

\begin{figure}[htbp]
\epsfig{file=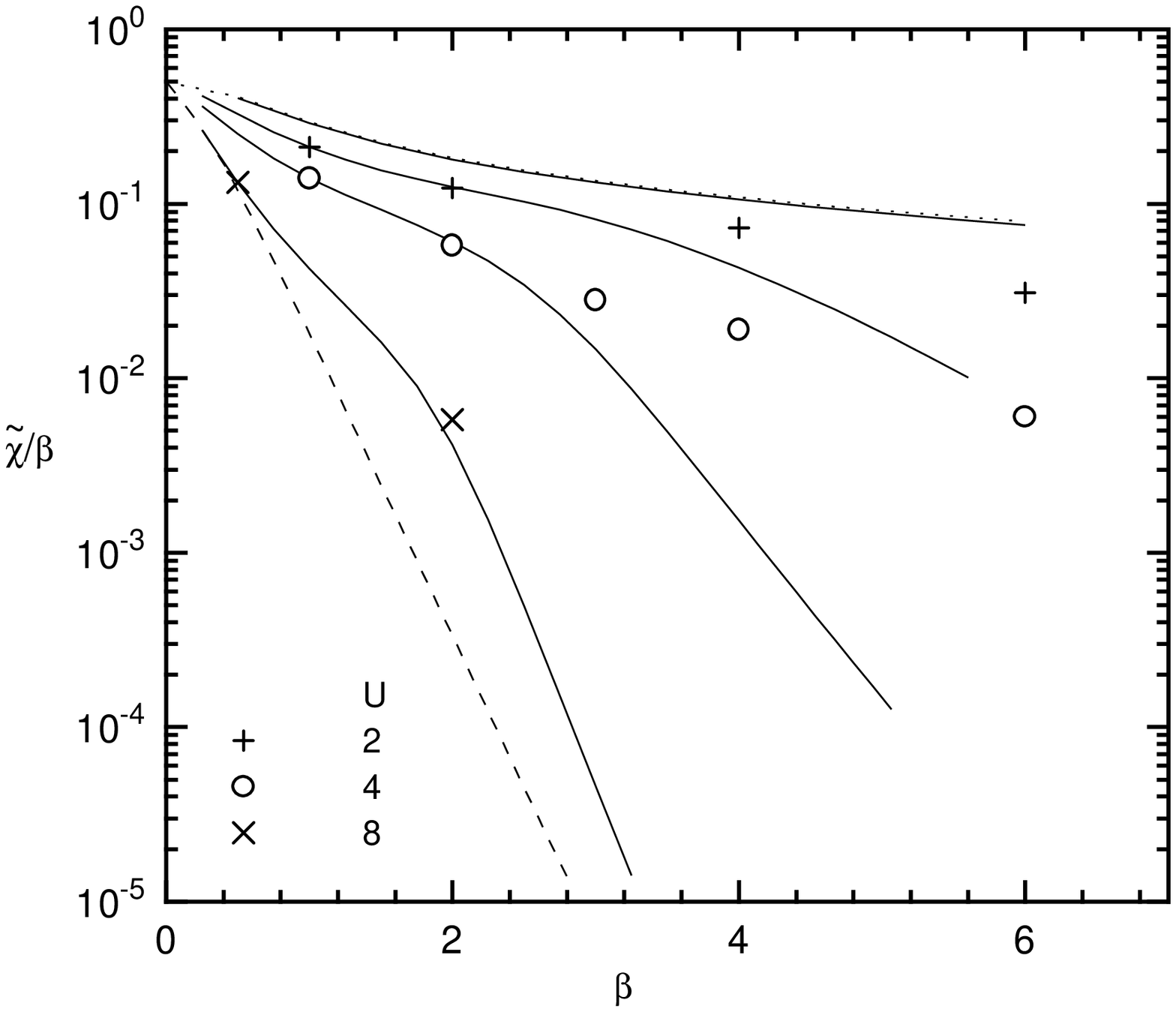, width=130mm}
\caption{ $\widetilde{\chi} / \beta$ as a function of $\beta$ for different
values of the coupling $U$ at $\mu=0$. See description in the text.}
\label{figxy}
\end{figure}

In the absence of any information on the analytic structure of
$ \widetilde{\chi}$ in $k$ as the interaction is turned on, we
decided to perform the same change of variable (\ref{defu}) in all
cases and constructed the Pad\'e approximants to the series in $u$.
We observe that the [2,3] and [3,2] approximants systematically
lead to very similar results. Our results for $\mu=0$ are collected
in Fig. (\ref{figxy}), and compared with those of the numerical simulation,
as obtained with the largest lattices which were simulated for
each set of $\beta,U$ values. We now describe and comment this
figure.

The upper (dotted) curve is the exact result in the
free case $U=0$. The four continuous lines are the [2,3] Pad\'e
results for respectively $U=0.1, 2, 4, 8$,
from top to bottom . The lower (dashed)
line is the ``atomic'' limit $(k=0)$ drawn for $U=8$. The
statistical errors are smaller than  the size of the symbols
representing our numerical results for the largest available
values of $N_s$, $N_t$. Tables (\ref{Res4},\ref{Res28}, \ref{compNt})
show that sizable finite size effects are present.
In particular, unlike $\widetilde{P}_0$, the susceptibility appears
to be rather sensitive to the number $N_t$ of time slices.
When comparisons can be made, our
results are in general agreement with published data \cite{RT,Morscal,SP}.

The $U=0.1$ curve is used as an overall check that the whole
series constructed from \cite{OH} is correctly implemented 
and that the change of variable and Pad\'e reconstruction
work close to the free case. For $U=2$, excellent agreement
between series and data is obtained at $\beta=1$ and $2$. The
figure suggests that at $\beta=3$, the series still gives a 
reasonable answer, although lower than the true one by 
probably 10 to 15 \%. A
similar departure shows up above $\beta=2$ at $U=4$, and, say,
$\beta=1.5$ at $U=8$.
Hence the overall pattern is that our treatment of the series gives
good results at $\beta$ values below a maximum which decreases as 
$U$ increases. Above this maximum, the Pad\'e approximant drops down
rapidly the more so when $U$ is large.
It is interesting to discuss this latter feature. Consider again
the case $U=8$. The dashed line $(k=0)$ is the curve (see appendix A)
\begin{equation}
\frac{\widetilde{\chi}}{\beta} = \frac{w}{1 + w}
\mbox{ , } w = \exp\left[ - \beta U /2  \right]
\hspace{12pt},
\end{equation}
drawn for $U=8$, thus falling down like $\exp \left[ -4 \beta \right]$.
One observes that the Pad\'e result at $U=8$ drops down roughly
parallel to it above $\beta=2$. 

Inspection of the series for $\widetilde{\chi}$ reveals that all (known)
terms of its expansion in $k^2$ contain at least one power of $w$ as
a factor. So any truncation unavoidably leads to an exponential fall
off in $\beta$ at fixed $U$, the more so when $U$ is large. This is 
what we observe with the series at hand, while the numerical
simulation indicates that, although  $\widetilde{\chi}$ actually
decreases as $\beta$ increases, it does so at a {\it
much slower} rate than naively expected. In other
words, there are collective effects which rather tend to maintain
a sizable susceptibility, even though low temperature or/and
large $U$ favor the formation of local pairs, which contribute zero
to the total spin. We note that, as already mentioned in the previous
section, the probability for single occupation, $S_1$ (as given in
Tables (\ref{Res4},\ref{Res28})) and thus the average local pair number 
$\left< n /2 \right> - S_1/2$, depends weakly upon temperature.
Hence the decrease of  $\widetilde{\chi}$ with $T$ is
{\it not } a consequence (only) of
pair formation at fixed $U$ and $\left< n \right>$. Its
sensitivity to $U$ also is strong compared to that of $S_1$.
As recently discussed in \cite{SP}, these questions
are of interest in connection with the kind of regime
(BCS-like or pair condensation) eventually leading to a 
superconducting transition. See also \cite{Morscal,RT}. 

Our results show that, unlike popular approximations such as RPA or T--matrix
treatment, the series expansion in $k^2$ can be continued successfully down
to fairly small $T$ values even at large coupling.
In order to further illustrate this point, we present our predictions for
$\widetilde{\chi}$ as a function of $T$ for $U$ = 4, 6, 8 and 12, at half
filling ($\left< n \right> = 1$), in Fig. (6).
These predictions appear as solid lines when our calculation is accurate
(the minimum $T$'s of the solid lines are conservative guesses).
We point out that by themselves these results clearly expose the main features
of physical interest, namely the existence of a crossover from a smooth regime
at high $T$ to a rapid fall off below a $U$ dependent $T$ value. Furthermore,
we observe that these $\widetilde{\chi}$ values are quantitatively very close
to that of \cite{SP} obtained numerically at the same $U$ values, but for
$\left< n \right> = 0.8$.
This enforces our general statement that there is no evidence for any
substantial dependence on filling.
%
%
\section*{7. Summary and conclusion}
%
%
Motivated by the interest for strongly interacting systems
of electrons in the context of high $T_c$ materials, we
have performed an investigation of the attractive Hubbard model.

Based on the hybrid Monte-Carlo method to simulate the path integral
representation of the partition function, a numerical study
provided us with estimates of the static zero-momentum
pair correlation $\widetilde{P}_0$ and spin susceptibility $\widetilde{\chi}$.
The exploration
covered the ranges [0.5, 6.0] for the inverse temperature $\beta$ and
[2.0, 8.0] for the interaction strength $U$ in units of the hopping
parameter $k$. At $U$=4 we investigate in detail the dependence on the chemical
potential in order to study the sensitivity of these physical quantities to
band filling, at and away from half filling.
We also performed an analytical investigation of the susceptibility,
based on its expansion in $k^2$ to order 5 derived from the results of
\cite{HOA} and on a specific Pad\'e resummation technique inspired by the
free case. We show that this method yields very good results
for $\beta \lesssim 1$ at any $U$, and $\beta \lesssim 2$ for $U \lesssim 4$.

In the numerical approach, the cost in computer time
leads to the consideration of rather small lattices only, and also
limits the number of time slices involved. The limitation of the
analytical approach is obviously due the truncation of the $k^2$ expansion,
which furthermore is not available for the pair correlation. But it yields
estimates of $\widetilde{\chi}$ directly in the thermodynamical limit, so
that the two approaches are complementary and their comparison brings
valuable information.
Whenever it was possible, we also compared our numerical data with
those of previous investigations \cite{Morscal,Scal,SP} performed using
Quantum Monte-Carlo. The overall agreement is good, although we
found significant discrepancies for $\widetilde{P}_0$ at our lowest T values.

Using pair correlations, the main question addressed was that of the
existence and location $T_{\rm KT}$ of a Kosterlitz-Thouless
superconducting transition, the only possible transition at finite
temperature away from half filling. Our results show very weak dependence
on $\mu$, hardly distinguishing half filling from other values of band filling
in the range between $.4$ and $.5$. Fits or finite size analysis relevant
to a KT transition are consistent with a transition at $T_{\rm KT}/k$ of
the order of a few percent, with a weak dependence on $\mu$, if any.
This conclusion is at variance with that drawn in \cite{Morscal}, but we
point out that it is perhaps consistent with what might follow from taking
into account the lowest temperature data points of this latter reference.
In any case, from our analysis we conclude that it is always necessary
to compare quantitatively the behavior at $\mu=0$, where there should not
be a transition at finite T to $\mu \neq 0$, before one can ascertain the
existence of a KT transition at finite T, and estimate its critical temperature.

Our study of the magnetic susceptibility shows that it depends
very weakly on the band filling, but decreases rapidly with increasing $U$
and/or decreasing temperature, in agreement with previous
investigations \cite{RT,Morscal,SP}. Increasing $U$ favors S-wave
pair formation, reducing the probability $S_1$ of sites occupied by a single
electron responding to a magnetic field. This explains the simultaneous
decrease of $\widetilde{\chi}$ and $S_1$ at fixed temperature. But at fixed $U$,
we observe that $S_1$ stays approximately constant in the low temperature
range whereas $\widetilde{\chi}$ falls steeply.
This casts some doubt on the existence of a tight connection between
susceptibility and pair formation as the expected transition is approached
as claimed in
\cite{SP}.
 
In this work, we restricted our data analysis to static, zero
momentum correlations. However, we have stored all the information
required to also analyze spatial correlations and
investigate dynamical properties of the model.

\section*{Acknowledgments}

We would like to thank Thierry Jolic\oe ur for many interesting discussions
and for his careful reading of our manuscript.
We further thank J. Oitmaa for providing us with the latest results on
the series expansion in suitable form.
Two of us (B. P. and J. S.) are grateful to the Deutsche Forschungsgemeinschaft
for support. Calculations were performed at HLRZ (J\"ulich) and CEA (Grenoble).
%
%
\section*{Appendix A}
%
%
We first discuss the atomic case $k=0$, at $h=0$
\begin{equation}
H = -U \left( a^{\dagger} a - \frac{1}{2} \right) 
\left( b^{\dagger} b - \frac{1}{2} \right)  - \mu
  \left(a^{\dagger} a+ b^{\dagger} b \right)
\hspace{12pt}.
\end{equation}

There are only 4 states. It is convenient to introduce the 
variables
\begin{equation}
x = e^{\beta \mu}
\hspace{12pt},
\end{equation}
\begin{equation}
w = e^{-\frac{\beta U}{2}}
\hspace{12pt}.
\end{equation}

One easily finds
\begin{equation}
Z = 1 + 2 w x + x^2
\hspace{12pt},
\end{equation}
\begin{equation}
\left< n \right> = \frac{2 x \left( w + x \right)}{1 + 2 w x + x^2}
\hspace{12pt},
\end{equation}
\begin{equation} \label{patom}
\widetilde{P}_0  = \frac{1+x^2}{ 1 + 2 w x + x^2}
\hspace{12pt},
\end{equation}
\begin{equation}
\widetilde{\chi} = \beta \frac{2 w x}{1 + 2 w x + x^2}
\hspace{12pt},
\end{equation}
\begin{equation}
\widetilde{C}(0,0) = \frac{2x \left( w + 2x \right) }{1 + 2 w x + x^2}
\hspace{12pt}.
\end{equation}

The single occupation probability is
\begin{equation}
 S_1 = \frac{2x  w}{1 + 2 w x + x^2} =\frac{\widetilde{\chi}}{\beta}
\hspace{12pt}.
\end{equation}

One remarks that $\widetilde{P}_0$ is always smaller than 1. The path integral 
expression of $Z$ factorises into one path integral for every site. It can
be analytically calculated, and with the form of the action which
we use, the result is exact independent of the value of $N_t$,
i.e. there are no finite $N_t$ corrections. 

In the free case $U=0$, the Hamiltonian, diagonalized in momentum space, is
\begin{equation}
H_{U=0}=\sum_p(\epsilon_p-\mu )
    (\tilde{a}_p^{\dagger}\tilde{a}_p + \tilde{b}_p^{\dagger}\tilde{b}_p)
- h (\tilde{a}_p^{\dagger}\tilde{a}_p - \tilde{b}_p^{\dagger}\tilde{b}_p)
\hspace{12pt},
\end{equation}
where
 $\epsilon_p=-2k(\cos p_1+\cos p_2)$ for momenta $p=(p_1,p_2)$ and $h$ is the
magnetic field introduced for completeness.

On a finite lattice $p_i=\frac{2\pi}{N_s}n_i$, $n_i=1,2,\cdots ,N_s$.
The correspondence with the thermodynamical limit $N_s\longrightarrow\infty$
is given by
\begin{equation}
\frac{1}{N_s^2}\sum_p\longleftrightarrow \int_0^{2\pi}
\frac{dp_1dp_2}{4\pi^2}
\hspace{12pt}.
\end{equation}
Setting
\begin{equation} 
s_p=\exp [-\beta (\epsilon_p-\mu)]
\hspace{12pt},
\end{equation}
one derives the partition function
\begin{equation} 
Z=\prod_p(1+s_p{\rm e}^{\beta h})(1+s_p{\rm e}^{-\beta h})
\hspace{12pt},
\end{equation} 
and the zero field quantities of interest :
\begin{equation}
\left< n \right> =\frac{2}{N_s^2}\sum_p\frac{s_p}{1+s_p}
\hspace{12pt},
\end{equation}
\begin{equation}
\widetilde{P}_0  = \frac{1}{N_s^2}\sum_p\frac{1+s_p^2}{(1+s_p)^2}
\hspace{12pt},
\end{equation} 
\begin{equation}
\widetilde{\chi} = \beta \frac{2}{N_s^2}\sum_p\frac{s_p}{(1+s_p)^2}
\hspace{12pt},
\end{equation}
\begin{equation}
\widetilde{C}(0,0) = \frac{2}{N_s^2}\sum_p\frac{s_p(1+2 s_p)}{(1+s_p)^2}
\hspace{12pt}.
\end{equation}

The single occupation probability $S_1 = \left< n \right>-\left< n \right>^2/2$ is independent of
temperature at fixed filling. It varies from 0 at zero filling to
1/2 at half filling.
 
Using $s_p=1/s_{p-(\pi,\pi)}$ at $\mu=0$, the susceptibility at half-filling
becomes 
\begin{equation} 
\left. \frac{\widetilde{\chi}}{\beta}
\right|_{\mu=0}= \frac{1}{N_s^2}\sum_p\frac{1}{1+\cosh (\beta\epsilon_p)} 
\hspace{12pt}. 
\end{equation} 

%
%
\section*{Appendix B}
%
%
In order to include terms of order $1/N_t^2$ within the action,
we apply a Trotter-splitting  in the following way:
\begin{equation} \label{tro3}
e^{- \frac{\beta}{N_t} \left( H_K + H_V \right) }  = 
 e^{- \frac{\beta}{2N_t} H_K} e^{- \frac{\beta}{N_t} H_V}
     e^{- \frac{\beta}{2N_t} H_K} + O \left( \frac{1}{N_t^3} \right)
\hspace{12pt},
\end{equation}
where
\begin{equation}
H_K =- k \sum_{\left< x,y \right>}
    \left( a^{\dagger}_{x} a_{y} +  a^{\dagger}_{y} a_{x} 
    +  b^{\dagger}_{x} b_{y} +  b^{\dagger}_{y} b_{x} \right)
\hspace{6pt},
\end{equation}
and
\begin{equation}
H_V= - U \sum_x \left( a^{\dagger}_{x} a_{x}
    -\frac{1}{2} \right) \left( b^{\dagger}_x b_{x} -\frac{1}{2} \right)
 - \mu \sum_{ x } \left( a^{\dagger}_{x} a_{x} + b^{\dagger}_{x} b_{x} \right)
\hspace{6pt}.
\end{equation}
We then, as before, introduce an
auxiliary scalar field. Then we expand the three exponentials in (\ref{tro3}),
collect terms including order $1/N_t^2$ and write the summands of operators in
normal order.  In the resulting expression we can substitute the operators
by Grassmann variables. The procedure described gives rise to additional
terms in the action.
The spatial elements of the matrix in Eq. (\ref{matrix1}) now read
\begin{eqnarray} \label{matrix2}
 M_{x,t;y,t} &=& \frac{k \beta}{2 N_t} \sum\limits_{\hat{\nu}=
 \hat{1}}^{\hat{2}} \left( \delta_{y,x+\hat{\nu}} + \delta_{y,x- \hat{\nu}}
 \right) \times \nonumber \\
 \times &&\hspace{-12pt} \left\{ \exp
 \left[ \sqrt{ \frac{U \beta}{N_t} } \sigma_{x,t} 
-(U- \mu) \frac{\beta}{N_t} \right]  
+ \exp
 \left[ \sqrt{ \frac{U \beta}{N_t} } \sigma_{y,t} 
-(U- \mu) \frac{\beta}{N_t} \right]  \right\} \nonumber \\
 &+& \delta_{x,y} \left\{  \exp
 \left[ \sqrt{ \frac{U \beta}{N_t} } \sigma_{x,t} 
-(U- \mu) \frac{\beta}{N_t} \right] -1   \right\}  \nonumber \\
 &+&  \frac{k^2 \beta^2}{2 N_t^2} 
 \sum\limits_{\hat{\nu},\hat{\nu}^{\prime} = \hat{1}}^{\hat{2}}
 \left\{ \delta_{y,x+\hat{\nu} + \hat{\nu}^{\prime}} +
 \delta_{y,x-\hat{\nu} - \hat{\nu}^{\prime}} +
 2\delta_{y,x+\hat{\nu} - \hat{\nu}^{\prime}} \right\} 
\hspace{12pt}.
\end{eqnarray}
%
%
%
%

\begin{table}
\begin{center}
\begin{tabular}{||c||c|c|r||c|c|r||r|r||}
\hline 
Lattice&$U$&$\beta$&$-\mu$&$\tau$&$n_{\rm MD}$&Traj.&$\xi_{\sigma^{\prime}}$&Acc.\\
\hline 
$4^2 \times 16$ & 4 & 1 & .15 & 1.6 & 16 & 19270 &  37 & .99\\
$4^2 \times  8$ & 4 & 1 & .15 & 1.6 & 16 & 20000 &  23 & .99\\
$6^2 \times  8$ & 4 & 1 & .15 & 1.6 & 16 & 20000 &  21 & .98\\
$4^2 \times 16$ & 4 & 1 & .0  & 1.6 & 16 & 19160 &  38 & .99\\
$4^2 \times  8$ & 4 & 1 & .0  & 1.6 & 16 & 20000 &  23 & .99\\
$6^2 \times 16$ & 4 & 1 & .0  & 1.6 & 16 & 20000 &  36 & .98\\
$6^2 \times  8$ & 4 & 1 & .0  & 1.6 & 16 & 20000 &  22 & .98\\
$4^2 \times 16$ & 4 & 2 & .15 & 1.6 & 32 & 12500 &  33 & .98\\
$4^2 \times 16$ & 4 & 2 & .0  & 1.6 & 32 & 12500 &  33 & .98\\
$4^2 \times 24$ & 4 & 3 & .15 & 1.6 & 48 & 12250 &  81 & .95\\
$4^2 \times 24$ & 4 & 3 & .0  & 1.6 & 48 & 12500 &  54 & .95\\
$4^2 \times 32$ & 4 & 4 & .6  & 0.8 & 64 & 22500 &  91 & .99\\
$4^2 \times 32$ & 4 & 4 & .45 & 0.8 & 64 & 22210 & 191 & .99\\
$6^2 \times 32$ & 4 & 4 & .45 & 1.6 & 64 & 20970 &  43 & .85\\
$6^2 \times 32$ & 4 & 4 & .2  & 1.6 & 64 & 43340 & 158 & .83\\
$4^2 \times 32$ & 4 & 4 & .15 & 0.8 & 64 & 19800 & 290 & .98\\
$6^2 \times 32$ & 4 & 4 & .15 & 1.6 & 64 & 20220 &  67 & .82\\
$8^2 \times 32$ & 4 & 4 & .15 & 1.6 & 64 & 18490 & 126 & .71\\ 
$6^2 \times 32$ & 4 & 4 & .1  & 1.6 & 64 & 20530 &  50 & .82\\
$4^2 \times 32$ & 4 & 4 & .0  & 0.8 & 64 & 19960 & 228 & .98\\
$4^2 \times 64$ & 4 & 4 & .0  & 1.6 & 64 & 20000 & 121 & .94\\
$6^2 \times 32$ & 4 & 4 & .0  & 1.6 & 64 & 20270 &  82 & .84\\ 
$4^2 \times 64$ & 4 & 6 & .45 & 1.2 & 96 & 10000 & 127 & .98\\
$6^2 \times 64$ & 4 & 6 & .45 & 1.2 & 96 &  9200 & 109 & .88\\
$4^2 \times 64$ & 4 & 6 & .3  & 1.2 & 96 & 12500 & 292 & .96\\
$6^2 \times 64$ & 4 & 6 & .3  & 1.2 & 96 & 12500 & 146 & .87\\
$6^2 \times 64$ & 4 & 6 & .23 & 1.2 & 96 & 12500 & 170 & .85\\
$4^2 \times 64$ & 4 & 6 & .15 & 1.2 & 96 & 19300 & 335 & .96\\
$6^2 \times 64$ & 4 & 6 & .15 & 1.2 & 96 & 12500 & 565 & .85\\
$8^2 \times 64$ & 4 & 6 & .15 & 1.2 & 96 &  8200 & 250 & .64\\
$4^2 \times 64$ & 4 & 6 & .08 & 1.2 & 96 & 20000 & 743 & .96\\
$4^2 \times 64$ & 4 & 6 & .0  & 1.2 & 96 & 12500 & 185 & .97\\
$6^2 \times 64$ & 4 & 6 & .0  & 1.2 & 96 & 10000 & 179 & .85\\
\hline
\end{tabular}
\\
\caption{ Parameters of the simulations performed at $U=4$.
Also given are the number of trajectories generated, the $\sigma^{\prime}$ field
autocorrelation time $\xi_{\sigma^{\prime}}$ and the acceptance.}
\label{algo4}
\end{center}
\end{table}

\begin{table}
\begin{center}
\begin{tabular}{||c||c|c|r||c|c|r||r|r||}
\hline 
Lattice&$U$&$\beta$&$-\mu$&$\tau$&$n_{\rm MD}$&Traj.&$\xi_{\sigma^{\prime}}$&Acc.\\
\hline 
$4^2 \times  8$ & 2 & 1 & .0  & 1.6 &  8 & 20000 &  13 &  .98\\
$4^2 \times 16$ & 2 & 2 & .0  & 1.6 & 16 &  9230 &  15 &  .98\\
$6^2 \times 32$ & 2 & 4 & .15 & 1.6 & 32 & 16140 &  18 &  .96\\
$6^2 \times 16$ & 2 & 4 & .15 & 1.6 & 32 & 47000 &  13 &  .90\\
$6^2 \times 32$ & 2 & 4 & .0  & 1.6 & 32 & 47180 &  20 &  .96\\
$6^2 \times 16$ & 2 & 4 & .0  & 1.6 & 32 & 38000 &  13 &  .90\\
$6^2 \times 24$ & 2 & 6 & .3  & 1.6 & 48 & 17780 &  13 &  .84\\
$6^2 \times 24$ & 2 & 6 & .15 & 1.6 & 48 & 16810 &  28 &  .79\\
$6^2 \times 24$ & 2 & 6 & .0  & 1.6 & 48 & 23420 &  24 &  .77\\ 
\hline
$6^2 \times 32$ & 8 & 0.5 & .15 & 1.6 & 16 & 20700 &  240  & .99\\
$6^2 \times 32$ & 8 & 0.5 & .0  & 1.6 & 16 & 20500 &  162  & .99\\
$6^2 \times 32$ & 8 & 2.0 & .15 & 1.6 & 64 & 12612 &  1237 & .95\\
$6^2 \times 32$ & 8 & 2.0 & .0  & 1.6 & 64 & 18855 &  782  & .94\\
\hline
\end{tabular}
\\
\caption{ Parameters of the simulations performed at $U=$ 2 and 8.
Also given are the number of trajectories generated, the $\sigma^{\prime}$ field
autocorrelation time $\xi_{\sigma^{\prime}}$ and the acceptance.}
\label{algo28}
\end{center}
\end{table}

\begin{table}
\begin{center}
\begin{tabular}{||c||c|c|c||l|l|l|l|l||}
\hline   
Lattice&$U$&$\beta$&$\mu$&order&$\left< \sigma^{\prime}\right>$
&$\left< n \right>/2$&$\ \ \widetilde{P}_0$&$\widetilde{\chi}\times 10$\\ \hline
$4^2 \times 16$&4&1& 0.& 1 &1.76(1)&0.468(2)&1.060(2)&1.406(3)\\
$4^2 \times 48$&4&1& 0.& 1 &1.94(2)&0.492(4)&1.060(3)&1.398(3)\\
\hline
$4^2 \times 8$ &4&1& 0.& 2 &2.05(2)&0.501(4)&1.071(4)&1.301(8)\\
$4^2 \times 16$&4&1& 0.& 2 &1.96(3)&0.490(5)&1.062(5)&1.390(9)\\
\hline
\hline
$4^2 \times 32$&4&4& 0.& 1 &1.34(2)&0.401(3)&2.20(3)&0.81(1)\\
$4^2 \times 64$&4&4& 0.& 1 &1.66(3)&0.451(6)&2.27(6)&0.79(2)\\
\hline
$4^2 \times 32$&4&4& 0.& 2 &2.11(4)&0.514(4)&2.37(9)&0.51(2)\\
$4^2 \times 64$&4&4& 0.& 2 &1.98(3)&0.495(5)&2.55(9)&0.71(2)\\
\hline
\end{tabular}
\\
\caption {Comparison of data taken with the fermionic matrix to order
1 and 2 in 1/$N_t$. See Eqs. (\ref{matrix1}) and (\ref{matrix2}) respectively.}
\label{compordre}
\end{center}
\end{table}

\begin{table}
\begin{center}
\begin{tabular}{||c||c|c|l||l|l|l|l|l||}
\hline 
Lattice&$U$&$\beta$&$-\mu$&$\left< \sigma^{\prime}\right>$&$\left< n\right>/2$&$\ \ \widetilde{P}_0$&$\widetilde{\chi}\times 10$ & $S_1$\\ \hline 
\hline
$4^2 \times 8$ &4&1& .15&1.90(2)&.463(4)&1.072(4)&1.289(8) & .250(2)\\
$4^2 \times 8$ &4&1& .0 &2.05(2)&.501(4)&1.071(4)&1.301(8) & .252(2)\\
$6^2 \times 8$ &4&1& .15&1.91(1)&.463(2)&1.079(7)&1.300(8) &.253(1)\\
$6^2 \times 8$ &4&1& .0 &2.03(2)&.496(3)&1.075(6)&1.313(8) &.255(1)\\
\hline
$4^2 \times 16$&4&2& .15&1.81(3)&.440(5)&1.64(2)&1.16(2) & .240(2) \\
$4^2 \times 16$&4&2& .0 &2.02(3)&.494(4)&1.63(3)&1.18(2) & .244(2) \\
\hline
$4^2 \times 24$&4&3& .15&1.81(3)&.435(3)&2.18(5)&0.85(3) & .234(2) \\
$4^2 \times 24$&4&3& .0 &2.10(3)&.509(6)&2.23(5)&0.87(2) & .242(2) \\
\hline
$4^2 \times 32$&4&4& .6 &1.31(2)&.309(1)&2.47(7)&0.45(2) & .226(4)\\
$4^2 \times 32$&4&4& .45&1.40(3)&.335(2)&2.6(1)&0.46(2) & .234(1)\\
$4^2 \times 32$&4&4& .15&1.87(6)&.45(1)&2.40(6)&0.55(2) & .234(4)\\
$4^2 \times 32$&4&4& .0 &2.11(4)&.514(4)&2.37(9)&0.51(2) &.230(4)\\
$6^2 \times 32$&4&4& .45&1.38(1)&.329(1)&3.0(1)&0.72(1) & .226(1)\\
$6^2 \times 32$&4&4& .2 &1.74(2)&.422(2)&2.87(5)&0.697(9) & .234(1)\\
$6^2 \times 32$&4&4& .15&1.83(2)&.443(2)&3.1(1)&0.68(1) & .232(2)\\
$6^2 \times 32$&4&4& .1 &1.92(1)&.468(2)&3.1(2)&0.71(1) &.234(1) \\
$6^2 \times 32$&4&4& .0 &2.06(2)&.501(3)&3.05(9)&0.75(1) &.239(1)\\ 
$8^2 \times 32$&4&4& .15&1.86(2)&.449(2)&2.85(8)&0.72(1) &\\ 
\hline
$4^2 \times 64$&4&6& .45&1.31(3)&.321(2)&2.38(8)&0.14(2) &  .246(6) \\
$4^2 \times 64$&4&6& .3 &1.41(4)&.344(2)&2.9(1)&0.22(2) & .242(6) \\
$4^2 \times 64$&4&6& .15&1.61(4)&.396(3)&3.3(2)&0.23(2) & .234(4) \\
$4^2 \times 64$&4&6& .08&1.86(6)&.460(7)&2.58(9)&0.18(1) &  .222(4) \\
$4^2 \times 64$&4&6& .0 &2.03(4)&.503(7)&2.57(9)&0.17(2) &  .236(6) \\
$6^2 \times 64$&4&6& .45&1.32(2)&.322(2)&4.0(3)&0.38(3) &  .224(6)  \\
$6^2 \times 64$&4&6& .3 &1.46(2)&.359(1)&4.2(2)&0.37(4) &  .240(4)  \\
$6^2 \times 64$&4&6& .23&1.62(2)&.396(1)&4.2(2)&0.42(2) &  .244(4)  \\
$6^2 \times 64$&4&6& .15&1.83(5)&.452(2)&3.9(3)&0.31(2) &  .238(4)  \\
$6^2 \times 64$&4&6& .0 &2.01(2)&.493(2)&3.8(2)&0.36(3) & .236(4) \\
$8^2 \times 64$&4&6& .15&1.75(3)&.429(1)&4.9(4)&0.49(3) &  .242(2) \\
\hline
\end{tabular}
\\
\caption {Data obtained on various lattices at $U=4$ :
$\left<\sigma^{\prime}\right>$, filling of the band $\left< n \right>/2 $, pairing correlation  $\widetilde{P}_{0}$,
spin susceptibility $\widetilde{\chi}$ and single occupation probability $S_1$
for different values of the inverse temperature
and of the chemical potential. }
\label{Res4}
\end{center}
\end{table}

\begin{table}
\begin{center}
\begin{tabular}{||c||c|c|l||l|l|l|l|l||}
\hline   
Lattice&$U$&$\beta$&$-\mu$&$\left< \sigma^{\prime}\right>$&$\left< n\right>/2$&$\ \ \widetilde{P}_0$&$\widetilde{\chi}\times 10$ & $S_1$\\ \hline
$4^2 \times  8$&2&1& .0 &1.45(1)&.502(2)&0.886(1)&2.114(4) &  .3738(8) \\
\hline
$4^2 \times 16$&2&2& .0 &1.42(2)&.497(3)&1.127(4)&2.46(1) & .373(1) \\
\hline
$6^2 \times 32$&2&4& .15&1.268(7)&.438(1)&1.46(1)&2.55(1) & .3732(8) \\
$6^2 \times 32$&2&4& .0 &1.453(4)&.504(1)&1.445(7)&2.727(9) & .3746(4) \\
\hline
$6^2 \times 24$&2&6& .3&1.153(4)&.3747(8)&1.99(2)&1.30(2) & .3446(6) \\
$6^2 \times 24$&2&6& .15&1.340(8)&.440(2)&2.20(6)&1.69(2) & .3444(8)\\
$6^2 \times 24$&2&6& .0 &1.605(6)&.531(2)&2.10(3)&1.86(2) & .346(1) \\ 
\hline
\hline
$6^2 \times 32$&8&.5& .15&2.69(7)&.475(8)&0.948(3)&0.667(5) & .174(2)\\
$6^2 \times 32$&8&.5& .0 &2.88(6)&.509(8)&0.955(4)&0.661(5) & .173(2) \\
\hline
$6^2 \times 32$&8&2.& .15&2.3(2)&.41(3)&1.9(1)&0.104(5) & .079(2) \\
$6^2 \times 32$&8&2.& .0 &2.7(1)&.47(2)&2.2(2)&0.116(5) & .085(2) \\
\hline
\end{tabular}
\\
\caption {Data obtained on various lattices at $U=$ 2 and 8. As in Table (\ref{Res4}).}
\label{Res28}
\end{center}
\end{table}

\begin{table}
\begin{center}
\begin{tabular}{||l|l l l|c|c||}
\hline
fitting function &  $b$ & $a$ & $T_{\rm KT}$ & $N_{pts}$& $ \chi^2 / d.o.f. $ \\ \hline 
$b \exp \left[ a / \sqrt{T-T_{\rm KT}} \right]$ &  0.42(2) & 0.92(6) & 0.03(2) & 5 & 1.58 \\
 &  0.46(9) & 0.9(2) & 0.04(3) & 4 & 2.96 \\
$ b \exp \left[ a / T \right]$ &  0.768(8) & 0.350(7) & & 5 & 10.55 \\
 &  0.90(3) & 0.30(1) & & 4 & 1.73 \\
$ b \exp \left[ a / \sqrt{T-0.1} \right]$ & 0.545(9) &0.66(1) &0.1 & 5 & 16.98 \\
& 0.68(3) &0.55(2) &0.1 & 4 & 6.02 \\
\hline 
\end{tabular}
\\
\caption{
Fits of  $\widetilde{P}_{0}$ as a function of $T$
for $U=4$ and $\mu = -0.15$. The point $T=1$ is ignored in every
second line. In the last two lines, $T_{\rm KT}$=0.1 [5]  
is imposed.
}
\label{fit}
\end{center}
\end{table}

\begin{table}
\begin{center}
\begin{tabular}{||c|l|l|l|l|l|l||} \hline
$ \kappa $ & 0 & 1 & 2 & 3 & 4 & 5 \\ \hline
$ \beta=1$ & 1. & .580 & -.725 & .372 & .156 &-.497 \\
$ \beta=2$ & 1. & 5.65 & 2.91 & -17.6 & -.404 & 62.5 \\ \hline 
\end{tabular}
\caption{ The ratios    $A_{\kappa} /     
A_{0}$ of the expansion of 
$\widetilde{\chi}$ in powers of $k^2$ for $U=4$, $\mu=0$, $\beta=1$
and 2.}
\label{x}
\end{center}
\end{table}

\begin{table}
\begin{center}
\begin{tabular}{||c||c|c|l||l|l|l|l|l||}
\hline   
Lattice&$U$&$\beta$&$-\mu$&$\left< \sigma^{\prime}\right>$&$\left< n\right>/2$&$\ \ \widetilde{P}_0$&$\widetilde{\chi}\times 10$ & $S_1$\\ \hline
$6^2 \times 32$&2&4& .15&1.268(7)&.438(1)&1.46(1)&2.55(1) & .3732(8) \\
$6^2 \times 16$&2&4& .15&1.377(4)&.4514(8)&1.67(1)&2.114(9) & .3510(6)\\
\hline
$6^2 \times 32$&2&4& .0 &1.453(4)&.504(1)&1.445(7)&2.727(9) & .3746(4) \\
$6^2 \times 16$&2&4& .0 &1.583(4)&.523(1)&1.65(1)&2.212(9) & .3532(6)\\
\hline
$4^2 \times 16$&4&1& .15&1.81(3)&.453(5)&1.062(5)&1.379(9) & .264(2) \\
$4^2 \times 8$ &4&1& .15&1.90(2)&.463(4)&1.072(4)&1.289(8) & .250(2)\\
\hline
$4^2 \times 16$&4&1& .0 &1.96(3)&.490(5)&1.062(5)&1.390(9) & .266(2)\\
$4^2 \times 8$ &4&1& .0 &2.05(2)&.501(4)&1.071(4)&1.301(8) & .252(2)\\
\hline
$6^2 \times 16$&4&1& .0 &2.02(2)&.501(3)&1.072(7)&1.382(8) &.267(1)\\
$6^2 \times 8$ &4&1& .0 &2.03(2)&.496(3)&1.075(6)&1.313(8) &.255(1)\\
\hline
$4^2 \times 32$&4&4& .0 &2.11(4)&.514(4)&2.37(9)&0.51(2) &.230(4)\\
$4^2 \times 64$&4&4& .0 &1.98(3)&.495(5)&2.55(9)&0.71(2) &.244(2)\\
\hline
\end{tabular}
\\
\caption {Comparison of data taken at different $N_t$ values.}
\label{compNt}
\end{center}
\end{table}

%
%
%
\newpage

\begin{figure}[htbp]
\epsfig{file=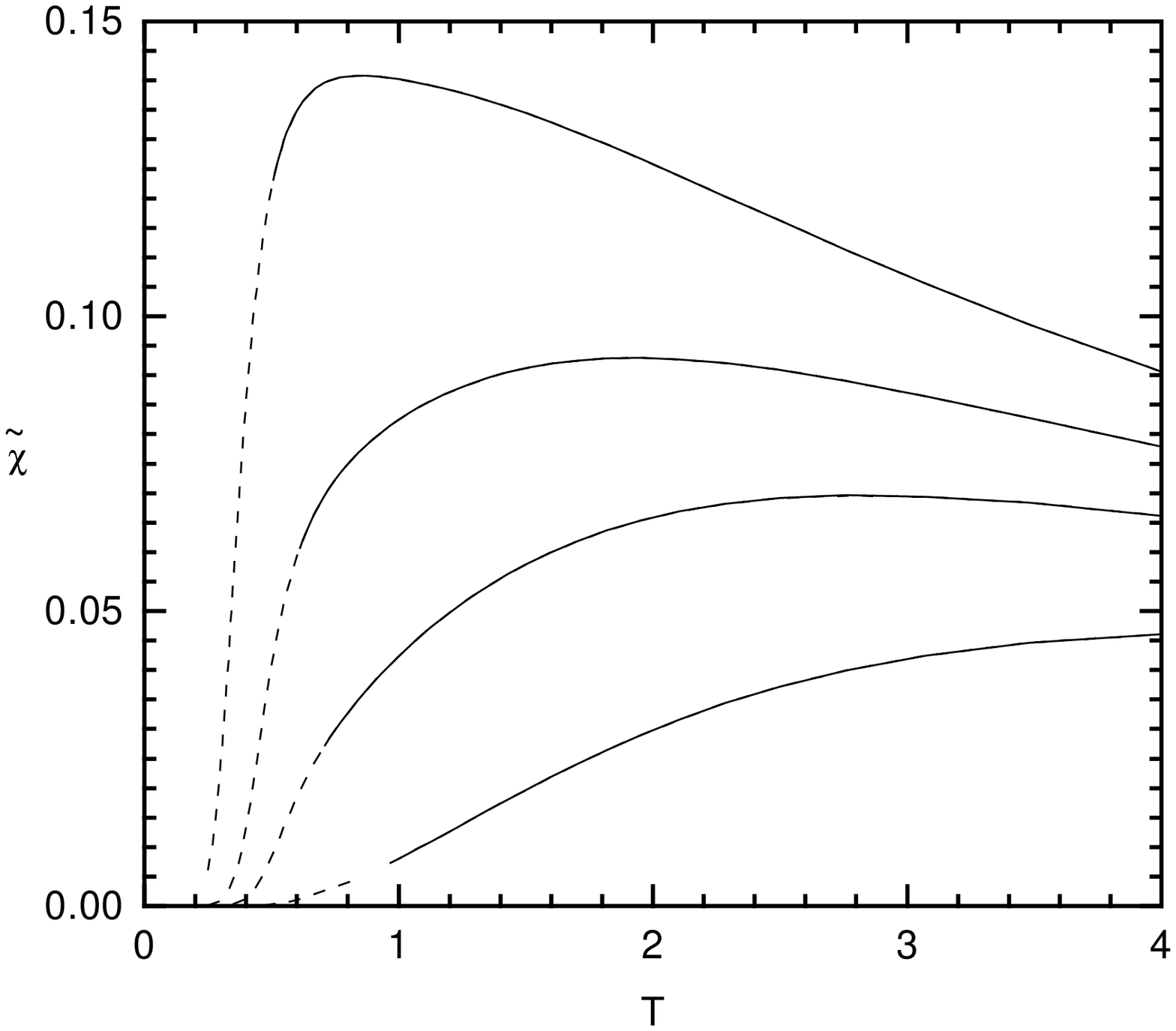, width=130mm}
\caption{ $\widetilde{\chi}$ as a function of $T$ for different values of the
coupling $U$ at $\mu=0$. From top to bottom  $U=4,6,8$ and $12$.
See description in the text.}
\label{figchi}
\end{figure}

\end{document}